\documentclass[]{aastex62}
\usepackage{amsmath}
\usepackage{bm}
\usepackage{amsthm}
\usepackage{mathrsfs}
\usepackage[outdir=./]{epstopdf}
\usepackage{url}

\DeclareMathOperator\erf{erf}

\received{November 27, 2021}
\accepted{January 5, 2022}
\submitjournal{\apj}

\shorttitle{Estimate of the Orphan-Chenab Stream's Progenitor}
\shortauthors{Mendelsohn et al.}

\begin{document}

\title{Estimate of the Mass and Radial Profile of the Orphan-Chenab Stream's Dwarf Galaxy Progenitor Using MilkyWay@home}

    \author{Eric J. Mendelsohn}
    \affiliation{Department of Physics, Applied Physics and Astronomy, Rensselaer Polytechnic Institute, 110 8$^{\rm th}$ St., Troy, NY 12180, USA}
    
    \author{Heidi Jo Newberg}
    \affiliation{Department of Physics, Applied Physics and Astronomy, Rensselaer Polytechnic Institute, 110 8$^{\rm th}$ St., Troy, NY 12180, USA}
    
    \author{Siddhartha Shelton}
	\affiliation{Department of Physics, Applied Physics and Astronomy, Rensselaer Polytechnic Institute, 110 8$^{\rm th}$ St., Troy, NY 12180, USA}
	
	\author{Lawrence M. Widrow}
	\affiliation{Department of Physics, Engineering Physics and Astronomy, Queen’s University, Canada}
	
    \author{Jeffery M. Thompson}
	\affiliation{Department of Physics, Applied Physics and Astronomy, Rensselaer Polytechnic Institute, 110 8$^{\rm th}$ St., Troy, NY 12180, USA}
	
	\author{Carl J. Grillmair}
	\affiliation{IPAC, California Institute of Technology, 1200 E California Blvd., Pasadena, CA 91125, USA}

\begin{abstract}
We fit the mass and radial profile of the Orphan-Chenab Stream's (OCS) dwarf galaxy progenitor by using turnoff stars in the Sloan Digital Sky Survey (SDSS) and the Dark Energy Camera (DEC) to constrain N-body simulations of the OCS progenitor falling into the Milky Way on the 1.5 PetaFLOPS MilkyWay@home distributed supercomputer. We infer the internal structure of the OCS's progenitor under the assumption that it was a spherically symmetric dwarf galaxy comprised of a stellar system embedded in an extended dark matter halo. We optimize the evolution time, the baryonic and dark matter scale radii, and the baryonic and dark matter masses of the progenitor using a differential evolution algorithm. The likelihood score for each set of parameters is determined by comparing the simulated tidal stream to the angular distribution of OCS stars observed in the sky. We fit the total mass of the OCS's progenitor to ($2.0\pm0.3$) $\times 10^7 M_\odot$ with a mass-to-light ratio of $\gamma=73.5\pm10.6$ and ($1.1\pm0.2$)$\times10^6M_{\odot}$ within 300 pc of its center. Within the progenitor's half-light radius, we estimate total a mass of ($4.0\pm1.0$)$\times10^5M_{\odot}$. We also fit the current sky position of the progenitor's remnant to be $(\alpha,\delta)=((166.0\pm0.9)^\circ,(-11.1\pm2.5)^\circ)$ and show that it is gravitationally unbound at the present time. The measured progenitor mass is on the low end of previous measurements, and if confirmed lowers the mass range of ultrafaint dwarf galaxies. Our optimization assumes a fixed Milky Way potential, OCS orbit, and radial profile for the progenitor, ignoring the impact of the Large Magellanic Cloud.
\end{abstract}


\section{Introduction}

A few dozen dwarf galaxies are known to orbit around the Milky Way. Over the course of billions of years, these galaxies tidally disrupt and stretch around the Milky Way into tidal streams. The positions and velocities of the stars that make up these streams therefore carry information about the Galaxy’s gravitational field. As such, these dwarf galaxies act as gravitational probes for determining the distribution of gravitating mass in the Milky Way \citep{ibata2001,johnston2002,koposov2010,newberg2010,SagChapter,bonaca2018,ibata2021}. For example, the path of the stream probes the Galactic potential, the transverse motion of the stream probes the time-dependence of the potential, and the width of the stream probes the internal structure of the original progenitor \citep{willett2010}. This idea of mapping the Galactic potential with stellar streams is not new. However, the focus has mostly been on using streams to constrain the gravitational potential of the Galaxy and hence the structure of its dark halo \citep{koposov2010,bovy2016,bonaca2018}. However, it is clear that the streams are also affected by the internal structure of their progenitors. Here, we fix the potential and use one well-known halo stream to probe the internal structure of its progenitor.

\cite{belokurov2006} and \cite{grillmair2006} independently discovered this stellar stream while examining the Sagittarius stream. Due to the lack of a visible progenitor, the stream was named the Orphan Stream. The southern portion of the stream was later named Chenab \citep{shipp2018} before it was discovered that both pieces of the stream resulted from the tidal disruption of the same dwarf galaxy. Therefore, we will refer to the stream as the Orphan-Chenab Stream (OCS). Preliminary measurements of the OCS found that it contained old, metal-poor stars possibly from the merger of a satellite galaxy with the Milky Way. \cite{belokurov2007} placed a lower bound of $~10^5M_{\odot}$ on the total mass of the OCS's missing progenitor by studying the interactions of the OCS with the High Velocity Clouds known as Complex A, the dwarf galaxy Ursa Major II (UMa II), and other globular clusters (Segue I, Ruprecht 106, and Palomar 1). However, it was later shown in \cite{sales2008} that the OCS was kinematically separate from both Complex A and UMa II. In 2010, \cite{newberg2010} calculated the orbit of the OCS and showed that a progenitor with a total mass of $~2.5\times10^6M_{\odot}$ could be used to fit the tidal stream rather well. This mass is roughly 100 times smaller than most measured Ultra-Faint Dwarf (UFD) galaxies, and other estimates of the OCS's progenitor suggest a total mass on the order of $10^8$ to $10^9 M_{\odot}$ \citep{fardal2019,hendel2018}.

The disparity in the progenitor mass estimates is interesting because the measured velocity dispersions of stars in ultrafaint galaxies lead many to conclude that dwarf spheroidal galaxies, including ultrafaint dwarf galaxies, have $10^7 M_\odot$ of mass enclosed within the central 300 pc, independent of the dwarf galaxy's luminosity \citep{mateo1993,gilmore2007,strigari2008}. A mass of a few times $10^6 M_\odot$ is less than ultrafaint dwarf galaxies (UFDs) are believed to have. But most measurements of dwarf galaxy masses are derived from velocity dispersions and the assumption that dwarf galaxies are in equilibrium.

While equilibrium is a reasonable assumption for dwarf spheroidal galaxies, it has been suggested that observations should be obtained to look for signs of tidal stretching as a check \citep{battaglia2013}. UFDs in particular are susceptible to errors in measurement from velocity dispersion due to their small number of bright stars, complications in measuring velocities due to the presence of binary stars, and the possibility that tidal forces could make the assumption of equilibrium invalid. \cite{martin2008} find that UFDs are elongated, and suggest tidal disruption is the ``least problematic" explanation. Objects that are nearly completely disrupted or close to apogalacticon could exhibit velocity dispersions that are systematically an order of magnitude or more higher than equilibrium values due to contamination from extra-tidal stars \citep{smith2013,blana2015}. Depending on the angle of observation, enforcing dynamical equilibrium on a dwarf spheroidal galaxy undergoing tidal disruption can either overestimate the mass when measuring dispersion along its major axis or underestimate it when measuring on a perpendicular axis \citep{lokas2010}. As an example, the kinematics of stars in Willman 1 are so far from Gaussian that one cannot even pretend that it is in dynamical equilibrium for the purpose of computing a mass-to-light ratio \citep{willman2011}. In addition, Triangulum II, previously thought to be the most dark-matter-dominated galaxy known \citep{kirby2015}, has been downgraded to possibly ``a star cluster or tidally stripped dwarf galaxy" because the originally measured velocity dispersion was calculated including a star that is now known to be in a binary system \citep{kirby2017,buttry2021}. More recently, a new ultra-faint structure named DELVE 2 was discovered whose mass-to-light ratio is sensitive to whether or not the system is undergoing tidal disruption with the Large Magellanic Cloud (LMC), and thus, whether it is classified as a globular cluster or a UFD \citep{cerny2020}. Clearly, the mass-to-light ratios of UFDs, thought to have the highest mass-to-light ratios of any known objects, are uncertain.

Due to their status as the most dark matter dominated objects in the cosmos, UFDs are popular targets for dark matter indirect detection experiments. The lack of  gamma ray signals from the centers of ultrafaint galaxies is being used to place upper limits on the properties of the as yet undetected dark matter particles \citep{abdallah2020}. Null results from dark matter searches in dwarf galaxies with Fermi LAT data have provided some of the strongest constraints on the dark matter annihilation cross section \citep{ackerman2015}; more recently, very weak excesses have been found in three ultrafaint dwarf galaxies \citep{albert2017}, but one is already found to be from a background source \citep{li2021}. But indirect detection experiments rely heavily on the estimate of the amount of dark matter above background that they are targeting to determine the detection limits. If we find that UFDs are less massive than previously assumed, the constraints these experiments put on dark matter particles would be modified.

Using the petaFLOPS-scale MilkyWay@home volunteer supercomputer, \cite{shelton2021} showed that in a perfect world it would be possible to determine the mass and radial profile of both the stars and the dark matter in a dwarf galaxy progenitor that fell into the Milky Way and was ripped apart into a tidal stream, using only the density distribution of stars in the tidal stream. In this context a perfect world means that the Milky Way potential is known and does not change with time, the orbit of the progenitor in the potential is known, and the dwarf galaxy is known to consist of a Plummer sphere distribution of stars embedded in a Plumber sphere distribution of dark matter, created so that the combination is in stable equilibrium. In this case, only the mass and Plummer radius of the two dwarf galaxy components and the evolution time of the simulation need to be fit.

This method of determining the progenitor satellite's properties from the tidal stream it produces does not rely the on assumption of dynamical equilibrium. We use MilkyWay@home to generate a large population of simulated dwarf galaxies with varying masses and shapes. Each dwarf galaxy is placed within a static Milky Way potential and evolved for a given amount of time to create a simulated tidal stream. This simulated tidal stream is then compared with the measured distribution of stars in the actual OCS and is assigned a likelihood that a dwarf galaxy with those simulated parameters produced the observed stream. We use differential evolution to evolve the dwarf galaxy parameters until the generated tidal stream closely matches the stellar data.

Our goal is to use MilkyWay@home to measure the mass and shape of the stars and dark matter in the progenitor galaxy that was tidally disrupted to become the OCS. In Section \ref{sec:data}, we describe the methods we implemented to obtain the stellar data used to constrain the OCS's progenitor. In Section \ref{sec:Nbody}, we outline the models and methods our N-body simulator uses to generate a tidal stream from a set of input progenitor parameters. Within the same section, we also define the likelihood score, the metric by which we quantitatively measure the similarity between the simulated tidal stream and the data. In Section \ref{sec:Optimize}, we explain the algorithm we use on MilkyWay@home to find the progenitor parameters that best fit the data. In Section \ref{sec:results}, we report the raw findings of our optimizations, and in Section \ref{sec:discuss}, we discuss the implications of our results.

\section{Stellar Data of the OCS} \label{sec:data}

To optimize our dwarf progenitor's parameters, we need accurate stream data to compare against our simulations. To this end, we extract the density of stars along the OCS as well as its width using actual data from the sky. We then parse it into a binned histogram which MilkyWay@home will compare with simulations. We use data from both the Sloan Digital Sky Survey \citep[SDSS;][]{SDSS_paper} and the Dark Energy Camera \citep[DEC;][]{DECcitation} to map the OCS. We use the same Lambda-Beta $(\Lambda,B)$ coordinate system defined in \cite{newberg2010} to follow the stream across the sky. Since the OCS does not follow a great circle across the sky and does not maintain the same distance from us as a function of $\Lambda$, we apply corrections to the unextincted magnitude in the g-band ($g_0$) and the $B$ coordinate. The equations for these corrected values are as follows:

    \begin{equation}\label{g_corr}
	   g_{corr} = g_{0} - 0.00022\Lambda^{2} + 0.034\Lambda ,\\
    \end{equation}
    
    \begin{equation}\label{B_corr}
	   B_{corr} = 
	   \begin{cases}
	   B + 0.00628\Lambda^{2} + 0.42\Lambda + 5.0 \indent &\Lambda\leq-15.0^{\circ}\\
	   B &\Lambda>-15.0^{\circ}
	   \end{cases},
    \end{equation}
as calculated in \cite{newberg2010}. To select F-turnoff stars at the same distance as the OCS, we filter stars such that $20.7<g_{corr}<21.7$. Because we removed faint stars (keeping $g_0<22.5$) so that we do not need to consider a variable reduction in completeness \citep[][]{Newberg2002,weiss_2018}, we miss fainter OCS stars at lower $\Lambda_{OCS}$, where the stream is further away. We correct for the missing data when generating the binned histogram.

\subsection{SDSS Data}\label{sec:sdss}

From the SDSS DR16 release \citep{SDSS_data}, we selected all stars with right ascension ($\alpha$) between and $123.75^\circ$ and $172.0^\circ$ and declination ($\delta$) between $-25.0^\circ$ and $60.0^\circ$.  We define our on-field to be the remaining stars within 2 degrees of $B_{corr}=0^\circ$ and our off-field to be the stars with $2.0^{\circ}<|B_{corr}|<4.0^{\circ}$. The difference between these two fields in a given $\Lambda$ bin is defined to be the excess in that bin. To avoid overlap between our two data sets, we also apply a $\Lambda$ cut to the SDSS data, keeping only stars with $\Lambda<21^\circ$.

We select F-turnoff stars with a $(g-i)_0$ color between 0.12 and 0.47. We select this color range to maximize the signal-to-noise ratio between the on-field and off-field stars (see Figure \ref{fig:gi_cuts}). We also filter stars such that $17.0<g_{0}<22.5$. To remove stars in front of or behind the OCS we enforced a distance cut of $20.7<g_{corr}<21.7$, which is the same cut applied in \cite{newberg2010}.

    \begin{center}
	\begin{figure}[!ht]
	    \centering
		\includegraphics[width = 9.5cm]{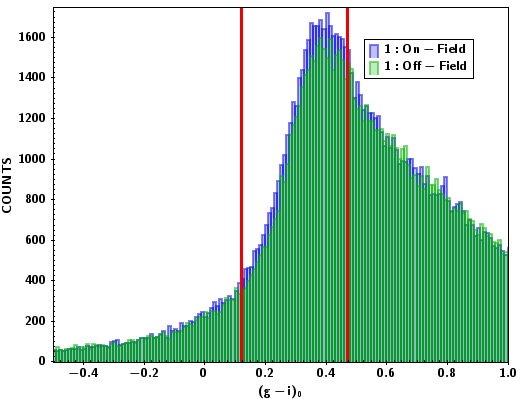}
	    \caption{Distribution of stars based on $(g-i)_0$ color. Blue represents stars in the on-field, and green reflects the off-field. The strongest signal can be detected by allowing stars with $0.12<(g-i)_0<0.47$, where we see the largest excess in on-field star counts. We represent the upper and lower bounds of this range with red lines in this plot. Only data from the SDSS is shown in this histogram.}
	\end{figure}\label{fig:gi_cuts}
    \end{center}
    
The errors in the counts are assumed to follow a Poisson distribution. Thus, the error of the counts in the $i^{th}$ $\Lambda$ bin $N_{i}$, for both the on-field and the off-field, is simply given by:

    \begin{equation}\label{N_err}
        \sigma_{N_{i}}=\sqrt{N_{i}}.
    \end{equation}

To account for the incompleteness in our fields at $\Lambda_{OCS}<-21^{\circ}$, we must calculate what fraction of each $\Lambda$ bin is filled. Since the incompleteness results from the $g_{0}=22.5$ magnitude limit, we can use \cite{newberg2010} to exactly determine the boundary $g_{b}(\Lambda_{OCS})$:

    \begin{equation}\label{eq:g_bound}
        g_{b}(\Lambda_{OCS})=a_g\Lambda_{OCS}^2 + b_g\Lambda_{OCS} + c_g,
    \end{equation}
where $a_g=-0.00022$, $b_g=0.034$, and $c_g=22.5$. As can be seen in Figure \ref{fig:g_corr_bin}, there are more F-turnoff stars at higher magnitudes in both the on-field and the off-field. We therefore correct for the missing stars by fitting a linear model to the star counts as a function of $g_{corr}$, as described in Appendix \ref{appendix:Incompleteness}.
    \begin{center}
	\begin{figure}[!ht]
	    \centering
		\includegraphics[width = 10.5cm]{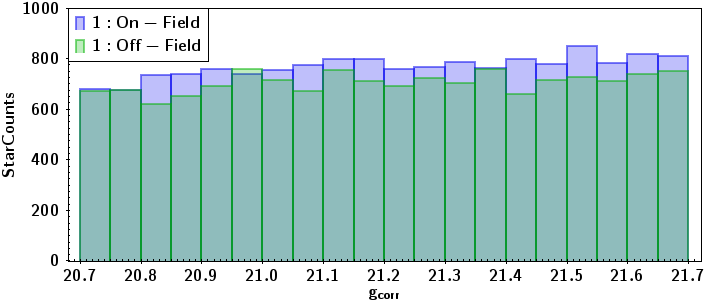}
	    \caption{F-turnoff stars from SDSS binned in $g_{corr}$ with $-21^{\circ}<\Lambda_{OCS}<21^{\circ}$, excluding the range $-7^{\circ}<\Lambda_{OCS}<10^{\circ}$ to avoid contamination from the Sagittarius Stream. We see a slight increase in star counts as we approach higher $g_{corr}$. The rate of increase is slightly higher for the on-field than for the off-field.}
	\end{figure}\label{fig:g_corr_bin}
    \end{center}

\subsection{DEC Data}\label{sec:dec}

The DEC data was taken from \cite{grillmair2015}. As the DEC uses the same g-band and i-band filters as the SDSS, we applied the same color and magnitude cuts as in the SDSS data, keeping stars with $21^{\circ}\leq\Lambda\leq48^{\circ}$. We define our on-field and off-field in the same way as we did for the SDSS data. However, as can be seen in Figure \ref{fig:Orphan_ONOFF}, the DEC data does not completely fill our on and off-fields.
    \begin{center}
	\begin{figure}[!ht]
	    \centering
		\includegraphics[width = 9.5cm]{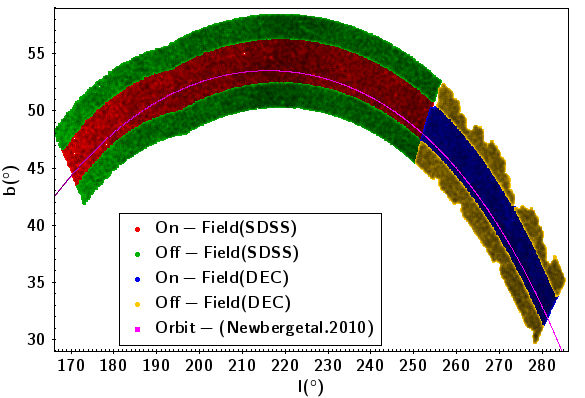}
	    \caption{Footprint of stellar data from the SDSS and DEC in Galactic longitude (l) and Galactic latitude (b). As can be seen in the DEC data (right), the data is incomplete as much of its off-field (yellow) is unobserved. There are also places where the on-field (blue) is incomplete. Our $(g-i)_0$ and $g_0$ cuts were used to generate this footprint. The pink line shows the orbit of the OCS as determined from \cite{newberg2010}.}
	\end{figure}\label{fig:Orphan_ONOFF}
    \end{center}

To correct our fields for the lack of data, we apply Monte Carlo approximations to ``fill in" the missing patches in the sky. For the $i^{th}$ $\Lambda$ bin in the on-field, we randomly populate the bin with 8192 ($2^{13}$) test stars. For each test star, we check whether the test star is within $0.18^{\circ}$ of a real star in that bin. We selected $0.18^{\circ}$ as our threshold because this was slightly larger than the maximum nearest-neighbor angular distance between two stars in the DEC data. We count the number of test stars within $0.18^{\circ}$ of a real star, $p_{i}$, and divide it by the total number of test stars, $M=8192$, to get the ``filled" fraction of the $i^{th}$ bin, $k_{i}$:

    \begin{equation}\label{k_ratio}
	   k_{i} = \frac{p_{i}}{M}.\\
    \end{equation}
Assuming a Poisson distribution for $p_i$, we derive the following expression for the error in $k_{i}$:

    \begin{equation}\label{k_err}
        \sigma_{k_{i}}=\frac{\sqrt{p_{i}}}{M}.\\
    \end{equation}
By dividing the number of counts in an on-field bin $N_{i}$ by the ratio $k_{i}$, we can approximate the true number of stars in that bin:

    \begin{equation}\label{N_prime}
        {N'}_{i}=\frac{N_{i}}{k_{i}}.\\
    \end{equation}
This means that the errors in counts in the $\Lambda$ bin are given by:

    \begin{equation}\label{eq:excess_errors}
        \sigma_{{N'}_{i}}= \frac{1}{{k_{i}}}\sqrt{ {\sigma_{N_{i}}}^{2} + {\left(\frac{N_{i}}{k_{i}}\right)}^{2} {\sigma_{k_{i}}}^{2}} = \frac{1}{{k_{i}}}\sqrt{ {N_{i}}\left(1 + \frac{{N_{i}}}{Mk_{i}}\right)}.
    \end{equation}
We apply the same process to each bin in the off-field as well.

\subsection{Distribution of Stream Stars along $\Lambda_{OCS}$}

After combining the stellar data from the SDSS with those from the DEC, we look at each bin and calculate the number of stars within the on-field and off-field. Using the off-field as a background, we subtract it from the on-field and determine the excess within the stream. The excess of the $i^{th}$ bin ($E_i$) and its error ($\sigma_{E_i}$) are calculated using the following formulae:

\begin{equation}\label{excess}
    E_{i} = 
    \begin{cases}
    N_{{\rm on},i}-N_{{\rm off},i} \indent &N_{{\rm on},i} \geq N_{{\rm off},i}\\
    0 &N_{{\rm on},i} < N_{{\rm off},i}\\
    \end{cases}
\end{equation}
\begin{equation}
    \sigma_{E_{i}}=\sqrt{ {\sigma_{N_{{\rm on},i}}}^{2}+{\sigma_{N_{{\rm off},i}}}^{2} }.
\end{equation}
The excess and errors of each bin are listed in Table \ref{tab:on_off_counts} along with the number of stars in each on-field and off-field bin. We also present a histogram representing this data in Figure \ref{fig:on_off_counts}. Using our cuts, we find an excess of $5,631\pm356$ F-turnoff stars in the OCS within the range of $-33^{\circ}\leq\Lambda\leq48^{\circ}$. For the purposes of our likelihood calculation which we will describe later, we also calculate the normalized excess star count ($e_i$) and its error ($\sigma_{e_i}$):

\begin{equation}
    e_i = \frac{E_i}{\sum_j E_j}
\end{equation}
\begin{equation}
    \sigma_{e_i} = \frac{1}{\sum_j E_j}\sqrt{(1-2e_i)\sigma_{E_i}^2+e_i^2\sum_j \sigma_{E_j}^2}
\end{equation}

    \begin{center}
	\begin{figure}[!ht]
	    \centering
		\includegraphics[width = 9.5cm]{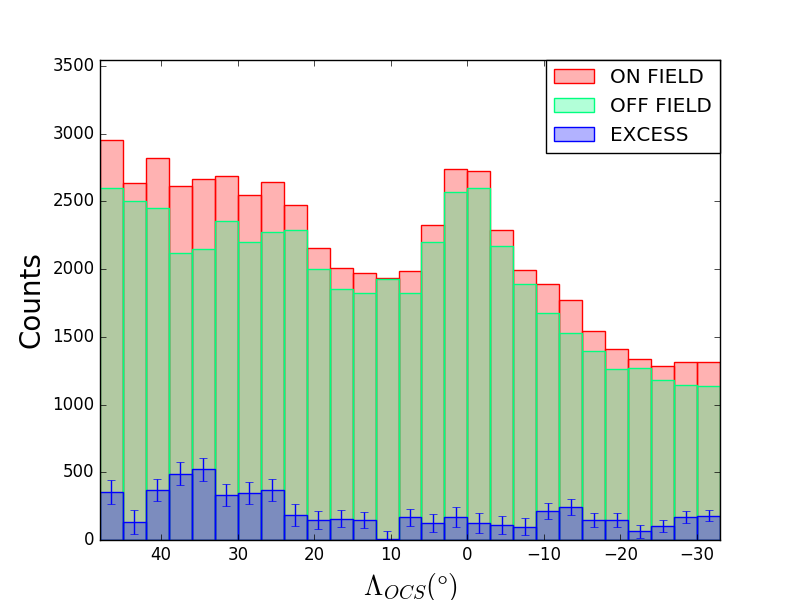}
	    \caption{The on-field (red), off-field (green), and excess(blue) of the OCS. Error bars of the excess are shown in the blue brackets.}
	\end{figure}\label{fig:on_off_counts}
    \end{center}
\begin{center}
    \begin{table}[!ht]
        \centering
        \begin{tabular}{ccccc}
             $\Lambda$ Center & On-Field & Off-Field & Excess ($E$) & Error ($\sigma_E$)\\
             \hline
             -31.5 & 1314 & 1136 & 178 & 40\\
             -28.5 & 1315 & 1144 & 171 & 44\\
             -25.5 & 1283 & 1178 & 105 & 46\\
             -22.5 & 1335 & 1271 & 64 & 49\\
             -19.5 & 1412 & 1262 & 150 & 52\\
             -16.5 & 1542 & 1395 & 147 & 54\\
             -13.5 & 1774 & 1530 & 244 & 57\\
             -10.5 & 1891 & 1676 & 215 & 60\\
             -7.5 & 1990 & 1893 & 97 & 62\\
             -4.5 & 2286 & 2173 & 113 & 67\\
             -1.5 & 2722 & 2596 & 126 & 73\\
             1.5 & 2737 & 2567 & 170 & 73\\
             4.5 & 2329 & 2201 & 128 & 67\\
             7.5 & 1989 & 1822 & 167 & 62\\
             10.5 & 1933 & 1927 & 6 & 62\\
             13.5 & 1968 & 1821 & 147 & 62\\
             16.5 & 2006 & 1850 & 156 & 62\\
             19.5 & 2152 & 2004 & 148 & 64\\
             \hline
             22.5 & 2480 & 2294 & 186 & 69\\
             25.5 & 2646 & 2269 & 337 & 70\\
             28.5 & 2546 & 2189 & 357 & 69\\
             31.5 & 2686 & 2356 & 330 & 71\\
             34.5 & 2668 & 2134 & 534 & 69\\
             37.5 & 2611 & 2163 & 448 & 69\\
             40.5 & 2817 & 2445 & 372 & 73\\
             43.5 & 2637 & 2491 & 146 & 72\\
             46.5 & 2953 & 2597 & 356 & 74\\
        \end{tabular}
        \caption{The number of F-turnoff stars in the on-field and off-field as a function of $\Lambda$. We notice a small gap in the OCS at $\Lambda=10.5^{\circ}$, however our optimizations will not be able to properly resolve the gap due to the large error of its associated bin. We split this table in terms of the SDSS data (top) and the DEC data (bottom).}
        \label{tab:on_off_counts}
    \end{table}
\end{center}
\subsection{Estimation of Total Stellar Mass}\label{sec:est_mass}
The simulations include bodies that represent both the baryons (stars) and the dark matter. The baryonic simulation bodies represent all of the stellar mass, not just the mass of the turnoff stars. Therefore, to constrain the mass of the OCS, we must relate the number of F-turnoff stars in the sky to an amount of baryonic mass. To accomplish this, we search for a globular cluster whose CMD properties most closely match those of the OCS. Once we find a globular cluster with a turnoff star color that is close enough to that of the OCS, we use that globular cluster to estimate the stellar mass per F-turnoff star in the OCS. Simply multiplying this ratio by the number of F-turnoff stars we detect in the OCS will give us an approximate stellar mass. For our analysis, we look at 11 candidate globular clusters using the globular cluster data from \cite{gc_data}.

It should be noted that because \cite{gc_data} uses the dust map from \cite{SFD1998} to calculate their unextincted magnitudes and our stellar data calculates extinctions using \cite{SF2011}, we need to correct the magnitudes from our globular clusters to adjust for over-reddening. Before we perform our globular cluster analysis, we add the extinction calculated from \cite{SFD1998} to each star's magnitude in \cite{gc_data} and then subtract the extinction derived from \cite{SF2011}.

When selecting globular cluster stars for our F-turnoff color calculation, we must search over a defined magnitude range for each cluster. To select this range, we plot a color-magnitude diagram (CMD) in $(g-i)_0$ and $g_0$. We set the minimum (brightest) turnoff magnitude at the intersection of the subgiants with the Red Giant Branch (RGB). We record the $(g-i)_0$ color of the intersection and determine the brightest magnitude for Main Sequence (MS) stars of that color, defining that $g_0$ as the maximum turnoff magnitude. The ranges are listed Table \ref{tab:gc_colors}, and the CMDs of each candidate can be seen in Figure \ref{fig:HR_clusters}.

Using these stars, we generate two one-dimensional histograms with bins in $(g-r)_{0}$ and $(g-i)_{0}$ color, respectively. We then fit the histogram to a skew normal distribution with a linear background. The model has the form:

\begin{equation}
    f(x)=mx+b+Ae^{-\frac{{(x-\xi)}^2}{2\omega^2}}\left(1+\erf{\left(\gamma \left(\frac{x-\xi}{\omega\sqrt{2}} \right) \right)}\right),
\end{equation}
where $m$ is the linear slope, $b$ is the linear y-intercept, $A$ is the amplitude, $\xi$ is the ``unskewed" mean, $\omega$ is the ``unskewed" standard deviation, and $\gamma$ is the skew parameter. In our fits, we assume that $\gamma\geq0$ because we expect the number of stars to drop off more slowly towards the redder end of the distribution. We fit these six parameters using a $\chi^2$ best-fit method and a differential evolution algorithm. We run the algorithm 10 independent times and keep the best fit to ensure our $\chi^2$ fit does not fall into a local minimum. After fitting the six parameters, we calculate the Hessian of the natural logarithm of the $\chi^2$ surface around that minimum. By inverting the Hessian, we derive the covariance matrix of the fitted parameters:

\begin{equation}\label{eq:Hessian}
    (H\ln{\chi^2})^{-1}=
    {\begin{pmatrix}
    \frac{\partial^2\ln{\chi^2}}{\partial{x_1}^2} & \frac{\partial^2\ln{\chi^2}}{\partial{x_1}\partial{x_2}} & \cdots\\
    \frac{\partial^2\ln{\chi^2}}{\partial{x_2}\partial{x_1}} & \frac{\partial^2\ln{\chi^2}}{\partial{x_2}^2} & \cdots\\
    \vdots & \vdots & \ddots
    \end{pmatrix}}^{-1}=
    \begin{pmatrix}
    {\sigma_{x_1}}^2 & \sigma_{x_1x_2}^2 & \cdots\\
    \sigma_{x_2x_1}^2 & {\sigma_{x_2}}^2 & \cdots\\
    \vdots & \vdots & \ddots
    \end{pmatrix}
\end{equation}

Each second derivative is calculated numerically using a finite step size, where each $i^{th}$ parameter is assigned a step size $h_i$. We approximate each second derivative using the following formula:

\begin{equation}
    \frac{\partial^2f}{\partial{x_i}\partial{x_j}} \simeq
    \begin{cases}
    \frac{f(x_i-h_i,x_j-h_j)-f(x_i-h_i,x_j+h_j)-f(x_i+h_i,x_j-h_j)+f(x_i+h_i,x_j+h_j)}{4{h_i}{h_j}} \indent& i\neq j\\
    \frac{f(x_i-h_i)-2f(x_i)+f(x_i+h_i)}{{h_i}^2} &i=j.\\
    \end{cases}
\end{equation}
After calculating the errors of each parameter, we take these errors and set them as the new step size for each of their respective parameters. We then calculate the Hessian again, taking those errors and feeding them back into the Hessian calculation over and over until the errors converge within a significantly small enough threshold (0.0001).

Taking our six fitted parameters and their errors, we calculate the mode of the fitted distribution. While skewed normal distributions do not have an analytical solution for the mode $Mo$, there does exist a numerical approximation for it while $\gamma\geq0$:

\begin{equation}
    Mo \simeq \xi + \omega\left(\zeta\sqrt{\frac{2}{\pi}} - \left(\frac{4-\pi}{4}\right)\frac{{\left(\zeta\sqrt{\frac{2}{\pi}}\right)}^3}{1-\frac{2}{\pi}\zeta^2} - \frac{1}{2}e^{-\frac{2\pi}{\gamma}}\right),
\end{equation}
where

\begin{equation}
    \zeta=\frac{\gamma}{\sqrt{1+\gamma^2}}.
\end{equation}

Running our SDSS OCS data through this pipeline, we find that the excess of the OCS has a peak $(g-r)_0$ color of $0.214\pm0.066$ and a $(g-i)_0$ color of $0.286\pm0.083$. We similarly run our 11 candidate globular clusters through our algorithm to determine which has the closest color peak. Table \ref{tab:gc_colors} lists the color peaks for each of these globular clusters.
\begin{center}
    \begin{table}[!ht]
        \centering
        \begin{tabular}{cccc}
             Globular Cluster & $g_0$ F-Turnoff Range & $(g-r)_0$ & $(g-i)_0$ \\
             \hline
             M2 & 19.0 - 21.0 & 0.351$\pm$0.015 & 0.457$\pm$0.028\\
             M3 & 18.5 - 20.5 & 0.309$\pm$0.022 & 0.382$\pm$0.047\\
             M5 & 18.0 - 20.0 & 0.326$\pm$0.024 & 0.456$\pm$0.051\\
             M13 & 18.0 - 20.0 & 0.318$\pm$0.035 & 0.425$\pm$0.056\\
             M15 & 18.75 - 20.75 & 0.338$\pm$0.020 & 0.483$\pm$0.033\\
             M53 & 19.75 - 21.5 & 0.264$\pm$0.034 & 0.351$\pm$0.042\\
             M92 & 18.0 - 20.0 & 0.262$\pm$0.027 & 0.359$\pm$0.043\\
             NGC 4147 & 20.0 - 21.5 & 0.267$\pm$0.038 & 0.353$\pm$0.050\\
             NGC 5053 & 19.5 - 21.5 & 0.241$\pm$0.025 & 0.326$\pm$0.041\\
             NGC 5466 & 19.25 - 21.25 & 0.283$\pm$0.024 & 0.351$\pm$0.029\\
             Palomar 5 & 20.5 - 22.25 & 0.361$\pm$0.040 & 0.505$\pm$0.053\\
        \end{tabular}
        \caption{Color peaks and the magnitude range for each of our 11 globular clusters.}
        \label{tab:gc_colors}
    \end{table}
\end{center}
    \begin{center}
	\begin{figure}[!ht]
	    \centering
		\includegraphics[width = 4cm]{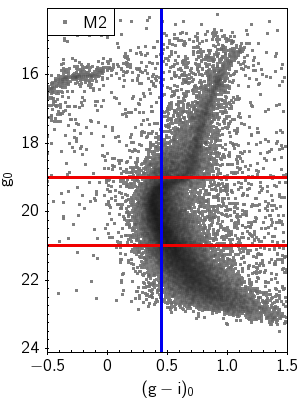}
		\includegraphics[width = 4cm]{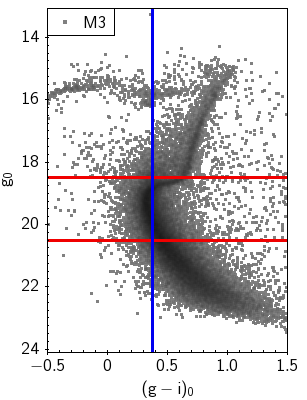}
		\includegraphics[width = 4cm]{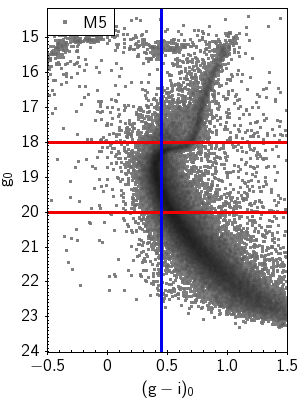}
		\includegraphics[width = 4cm]{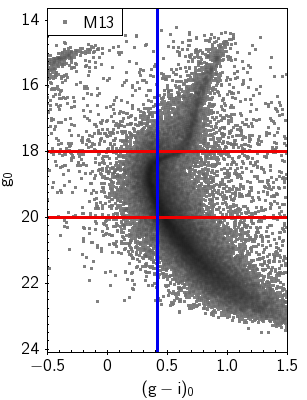}
		\includegraphics[width = 4cm]{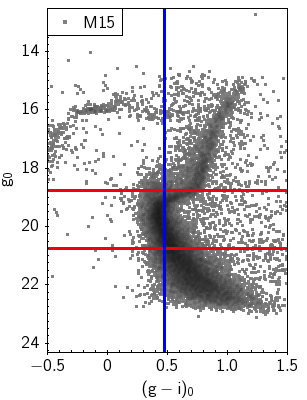}
		\includegraphics[width = 4cm]{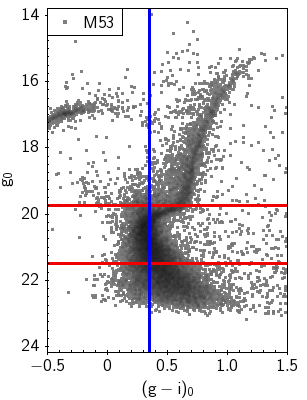}
		\includegraphics[width = 4cm]{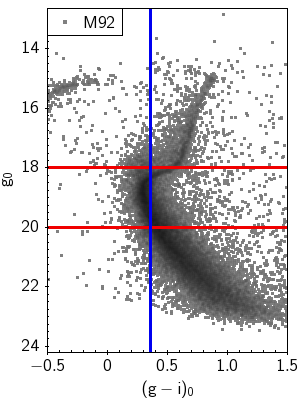}
		\includegraphics[width = 4cm]{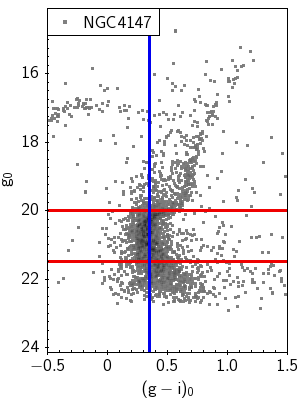}
		\includegraphics[width = 4cm]{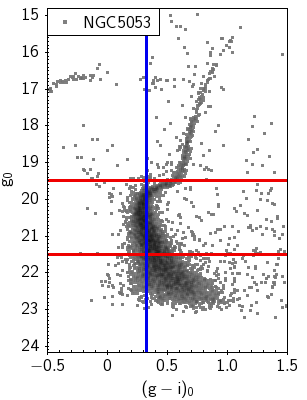}
		\includegraphics[width = 4cm]{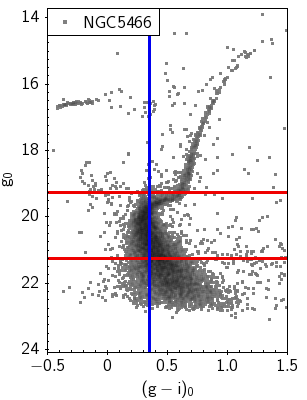}
		\includegraphics[width = 4cm]{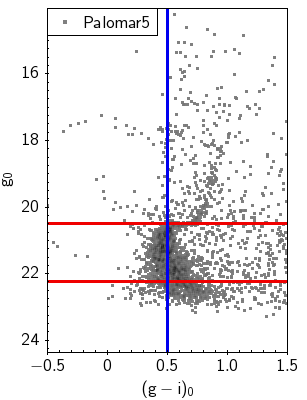}
	    \caption{CMDs of our 11 candidate globular clusters for our OCS color proxy. We plotted this data using TOPCAT \citep{TOPCATcitation} and globular cluster data from \cite{gc_data}. The red lines show the $g_0$ boundaries of our turnoff range, and the blue line represents the peak $(g-i)_0$ color.}
	\end{figure}\label{fig:HR_clusters}
    \end{center}
The globular cluster whose peak F-turnoff star color is most similar to that of the OCS is NGC 5053.

After determining which globular cluster to use, we  count the total number of F-turnoff stars we see in NGC 5053 from the photometric data provided by \cite{gc_data}, applying the same color cuts we employed in our stream separation, selecting stars with $0.12<(g-i)_0<0.47$ and $19.5<g_0<21.5$ as our turnoff stars. However, since globular clusters have a high stellar density in the sky, it is difficult to resolve all the stars within the center of the cluster. Although \cite{gc_data} uses a DAOPHOT pipeline that is specifically designed for crowded-field stellar photometry \citep{stetson1987} to parse the SDSS data, we find that there are still stars which are not resolved. To correct for the missing stars, we fit the radial distribution of NGC 5053 to that of a Plummer sphere. The 2D surface number density of a Plummer sphere is given by:

\begin{equation}
    \Sigma(R)=\frac{N_{total}a^2}{\pi{\left(R^2+a^2\right)}^2},
\end{equation}
where the number of stars ${\Delta}{N}$ within ${\Delta}{R}$ of radius $R$ from the center of the cluster is represented with the formula:

\begin{equation}
    {\Delta}{N}=2\pi{R}\Sigma(R){\Delta}{R}.
\end{equation}

We fit the total number of F-turnoff stars ($N_{total}$) and the angular scale radius ($a$) to NGC 5053 using the method of least squares, setting the radius $R$ for a given star as the angular distance between that star and the center of the cluster, which we find to be at $(\alpha,\delta)=(199.107^{\circ},17.6927^{\circ})$. To ensure our fit is not heavily impacted by the overcrowding near the center of the globular cluster, we exclude all stars within the central core by requiring $R>\theta_{1/2}\sim0.035^\circ$, where $\theta_{1/2}$ is the angular distance from the globular cluster's center where the stellar density drops to half of its maximum value. Figure \ref{fig:NGC5053_radial} shows the radial distribution of F-turnoff stars in NGC 5053 and the fitted Plummer curve.

    \begin{center}
	\begin{figure}[!ht]
	    \centering
		\includegraphics[width = 9cm]{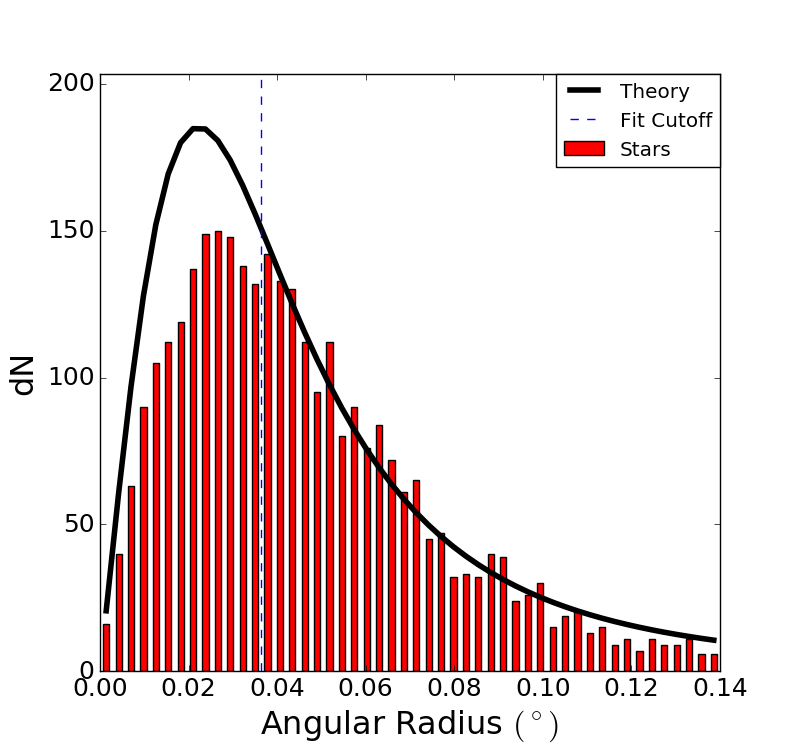}
	    \caption{Radial distribution of stars in NGC 5053 that passed our color cuts. The stellar densities (red) in the outer radii of the globular cluster fit relatively well to a Plummer distribution (black). Only data points to the right of the dotted blue line contributed to the fit. We notice a sizeable number of F-turnoff stars missing in the core due to high stellar density in the center.}
	\end{figure}\label{fig:NGC5053_radial}
    \end{center}

Correcting for missing stars near the center of the cluster, we find a total of $3,932\pm110$ F-turnoff stars within NGC 5053. Taking the mass of NGC 5053 to be $(5.37\pm1.32)\times10^4 M_{\odot}$ \citep{ngc5053_mass}, we calculate that each turnoff star represents a stellar mass of roughly $(13.7\pm3.4) M_{\odot}$. A calculation of the mass per turnoff star using isochrones, which finds a similar value, is done in Appendix \ref{appendix:Isochrone}. Multiplying this number by the 5,631$\pm$356 F-turnoff stars we calculate in the OCS gives us a total baryonic mass of $(7.7\pm2.0)\times10^4 M_{\odot}$ within the range $-33.0^{\circ}<\Lambda<48.0^{\circ}$.

\subsubsection{Systematic Error from Using a Redder Globular Cluster}\label{sec:ngc5053_error}

While the turnoff color of NGC 5053 matches that of the OCS within errors, the globular cluster does have a somewhat redder measure of turnoff color. To illustrate how a redder color would impact our estimate of the mass per turnoff star, we recalculate this quantity using all of our other globular clusters as the reference. Table \ref{tab:gc_compare} below shows the mass per F-turnoff star we find if we use each of these redder globular clusters as a reference.

\begin{center}
    \begin{table}[!ht]
        \centering
        \begin{tabular}{cccccc}
             Globular Cluster & $(g-r)_0$ & $(g-i)_0$ & Total Mass ($M_{\odot}$) & Num. F Stars & Mass/Turnoff Star ($M_{\odot}$) \\
             \hline
             M2 & 0.351$\pm$0.015 & 0.457$\pm$0.028 & (5.75$\pm$2.65)$\times10^5$ & 16,701$\pm$1,687 & 34.4$\pm$15.9\\
             M3 & 0.309$\pm$0.022 & 0.382$\pm$0.047 & (4.68$\pm$1.40)$\times10^5$ & 15,808$\pm$261 & 29.6$\pm$8.9\\
             M5 & 0.326$\pm$0.024 & 0.456$\pm$0.051 & (3.89$\pm$1.16)$\times10^5$ & 12,061$\pm$376 & 32.3$\pm$9.7\\
             M13 & 0.318$\pm$0.024 & 0.425$\pm$0.056 & (6.05$\pm$0.33)$\times10^5$ & 17,464$\pm$554 & 34.6$\pm$2.2\\
             M15 & 0.338$\pm$0.020 & 0.483$\pm$0.033 & (5.13$\pm$1.06)$\times10^5$ & 6,543$\pm$1,843 & 78.4$\pm$27.4\\
             M53 & 0.264$\pm$0.034 & 0.351$\pm$0.042 & (3.48$\pm$1.04)$\times10^5$ & 10,913$\pm$328 & 31.9$\pm$9.6\\
             M92 & 0.262$\pm$0.027 & 0.359$\pm$0.043 & (2.75$\pm$0.57)$\times10^5$ & 11,382$\pm$1,107 & 24.2$\pm$5.5\\
             NGC 4147 & 0.267$\pm$0.038 & 0.353$\pm$0.050 & (3.72$\pm$2.82)$\times10^4$ & 1,564$\pm$87 & 23.8$\pm$18.1\\
             NGC 5053 & 0.241$\pm$0.025 & 0.326$\pm$0.041 & (5.37$\pm$1.32)$\times10^4$ & 3,932$\pm$110 & 13.7$\pm$3.4\\
             NGC 5466 & 0.283$\pm$0.024 & 0.351$\pm$0.029 & (3.80$\pm$2.45)$\times10^4$ & 4,425$\pm$126 & 8.6$\pm$5.5\\
             Palomar 5 & 0.361$\pm$0.040 & 0.505$\pm$0.053 & (5.2$\pm$0.7)$\times10^3$ & 528$\pm$37 & 9.8$\pm$1.5\\
        \end{tabular}
        \caption{Mass per F-Turnoff star using different globular clusters as a reference. The total mass of each globular cluster, with the exception of M13, NGC 5053, and Palomar 5, came from \cite{kimmig2018}. M13's total mass was pulled from \cite{m13_mass} and Palomar 5's mass came from \cite{pal5_mass}.}
        \label{tab:gc_compare}
    \end{table}
\end{center}

Plotting the mass per turnoff star as a function of the peak color (Figure \ref{fig:color_v_turnoff}), we see that redder globular clusters typically produce a larger calculated mass per turnoff star. Fitting these points to an exponential curve of the form $\ln(M/M_{\odot})=r(g-i)_0+\ln(A)$, we find $r = 17.7\pm4.6$ and $\ln(A) = -3.96\pm1.89$ ($\chi^2=14.8$). This implies that using a globular cluster for the mass estimate of a stellar system with a bluer peak color overestimates its mass. Because we select a fixed (blue) color range to select our turnoff stars $(0.12 < (g-i)_0 < 0.47)$, it makes sense that using a redder globular cluster would result in fewer stars falling within this color range, and thus give us a larger stellar mass per turnoff star. If we extrapolate the curve to find the mass per turnoff star using a globular cluster with a color similar to the OCS, we find $3.0\pm6.9 M_{\odot}$ per turnoff star, roughly one quarter of the mass per turnoff star estimated using NGC 5053. It should be noted, however, that we do not use this extrapolated mass per turnoff for analyzing the OCS due the extrapolation's poor fit and large errors. We instead use the mass per turnoff derived from NGC5053, keeping in mind that any masses we fit using this data are overestimates.

    \begin{center}
	\begin{figure}[!ht]
	    \centering
		\includegraphics[width = 8.9cm]{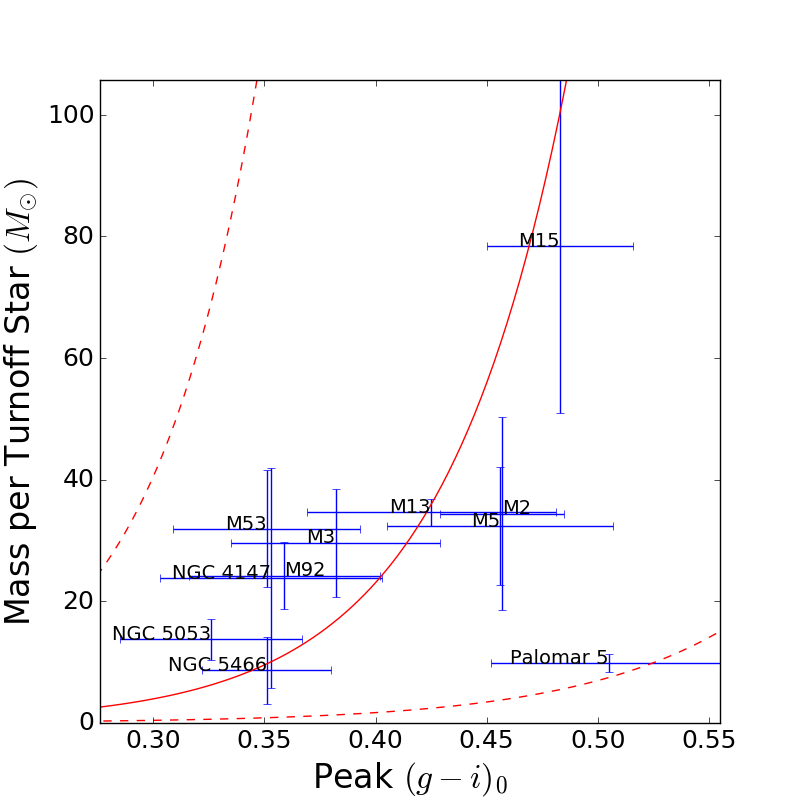}
	    \caption{A plot of the peak turnoff color versus the mass per turnoff star for various globular clusters. Redder globular clusters seem to have a larger mass per F-turnoff star, following an exponential curve. The solid red curve shows the curve of best fit ($\chi^2=14.8$) while the dashed red lines are the curves one standard deviation above and below the best fit. The curve has the form $\ln(M/M_{\odot})=r(g-i)_0+\ln(A)$, where $r = 17.7\pm4.6$ and $\ln(A) = -3.96\pm1.89$.}
	\end{figure}\label{fig:color_v_turnoff}
    \end{center}

\subsection{Beta Dispersion as a Function of Position along the Stream}

To fit the apparent width of the stream as a function of $\Lambda$, we measure the Beta dispersion of the OCS by splitting each $\Lambda$ bin into 20 $B$ bins, covering both the on and off fields. In each $\Lambda$ bin, we fit the distribution of $B$ values to a Gaussian distribution with a linear background, taking the fitted $\sigma$ value as the Beta dispersion:

\begin{equation}\label{eq:beta_model}
    f(x)=mx+b+Ae^{-\frac{{\left(x-\mu\right)}^2}{2\sigma^2}},
\end{equation}
where $m$ is the slope of the background, $b$ is the y-intercept, $A$ is the amplitude, $\mu$ is the mean, and $\sigma$ is the standard deviation of the Gaussian distribution.

To ensure a larger sample size, we group each consecutive set of three $\Lambda$ bins into a single bin for the purposes of the Beta dispersion calculation. We employ the same Monte Carlo method of bin completion used in Section \ref{sec:dec} to determine the number of stars in each Beta bin. However, we exclude any bins whose data coverage is less than half of the bin area from our fit. We use the ``curve fit'' algorithm from SciPy \citep{2020SciPy-NMeth} to fit the beta dispersions and their respective errors. Figure \ref{fig:beta_disp_fits} shows the fit in each set of $\Lambda$ bins while Table \ref{tab:beta_disp_fits} lists the numerical values of the Beta dispersion and their errors.

    \begin{center}
	\begin{figure}[!ht]
	    \centering
		\includegraphics[width = 5.9cm]{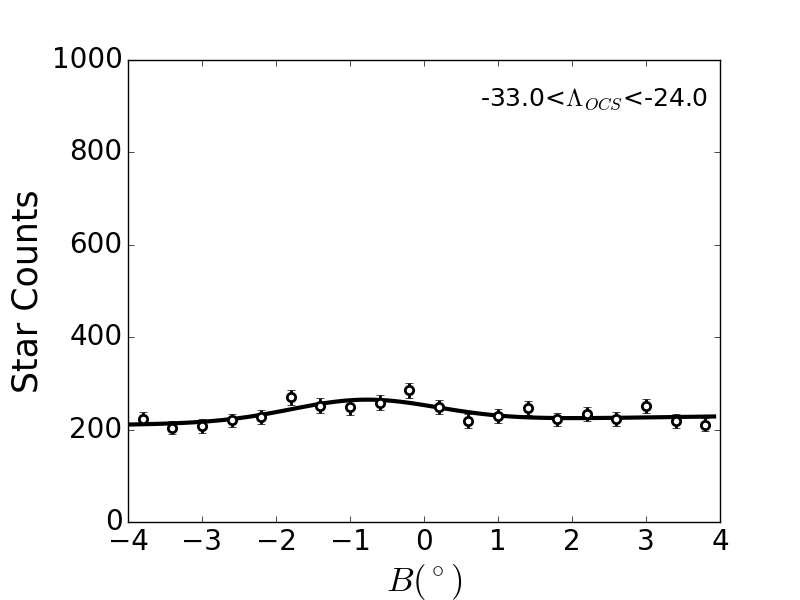}
		\includegraphics[width = 5.9cm]{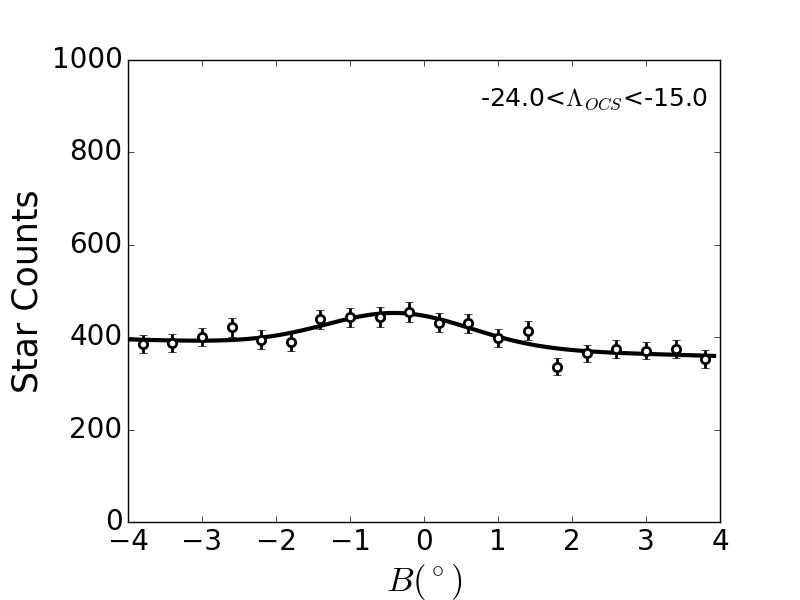}
		\includegraphics[width = 5.9cm]{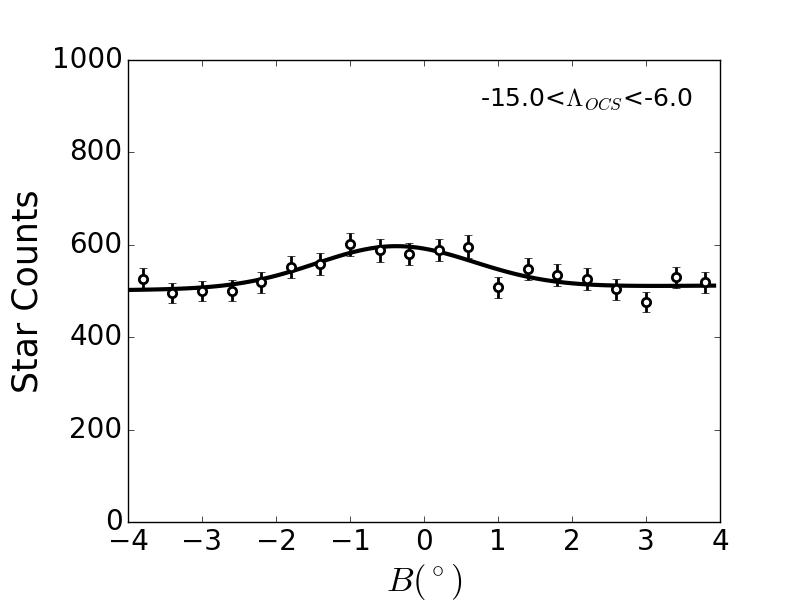}
		\includegraphics[width = 5.9cm]{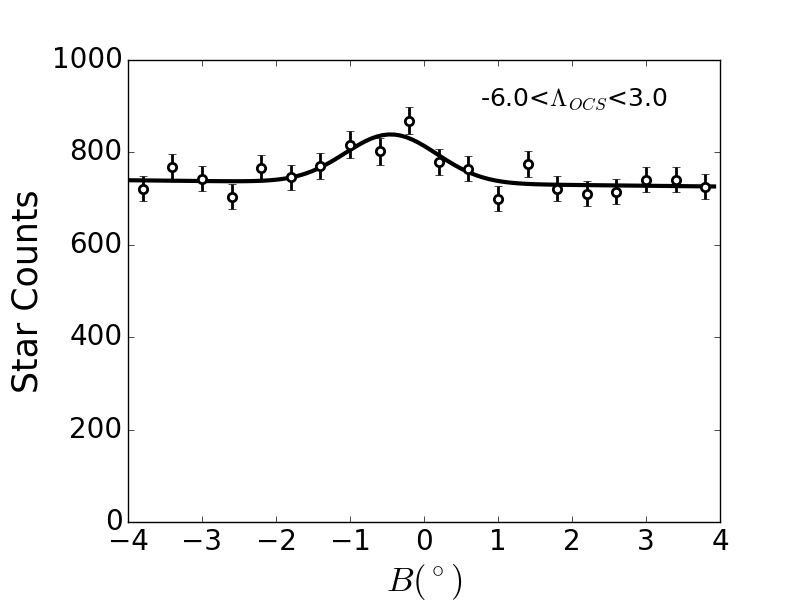}
		\includegraphics[width = 5.9cm]{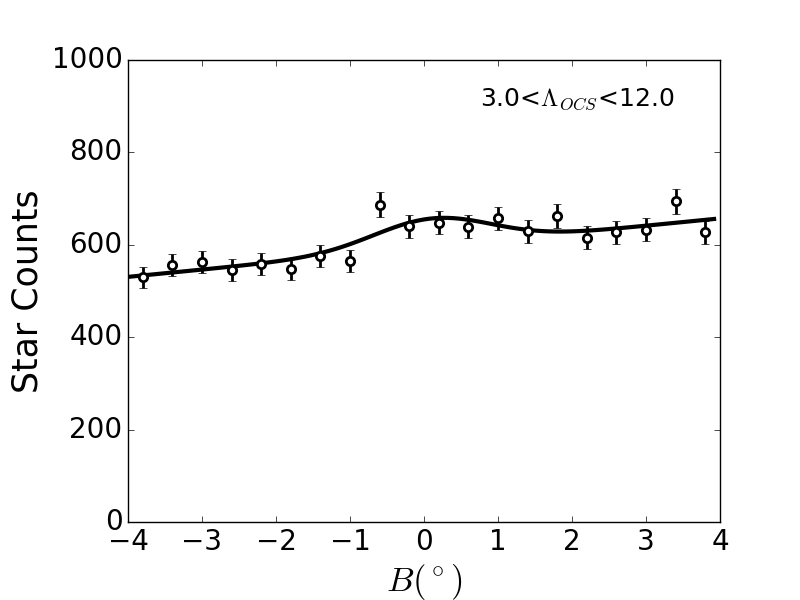}
		\includegraphics[width = 5.9cm]{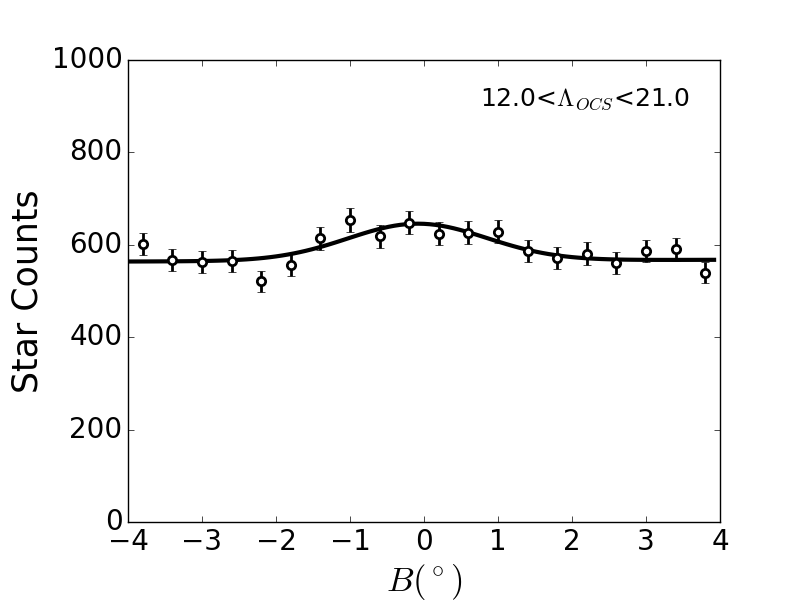}
		\includegraphics[width = 5.9cm]{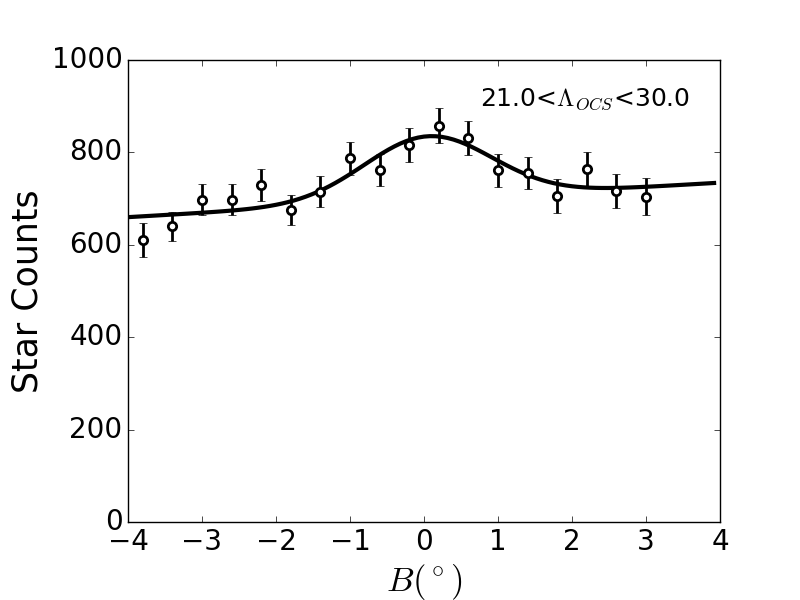}
		\includegraphics[width = 5.9cm]{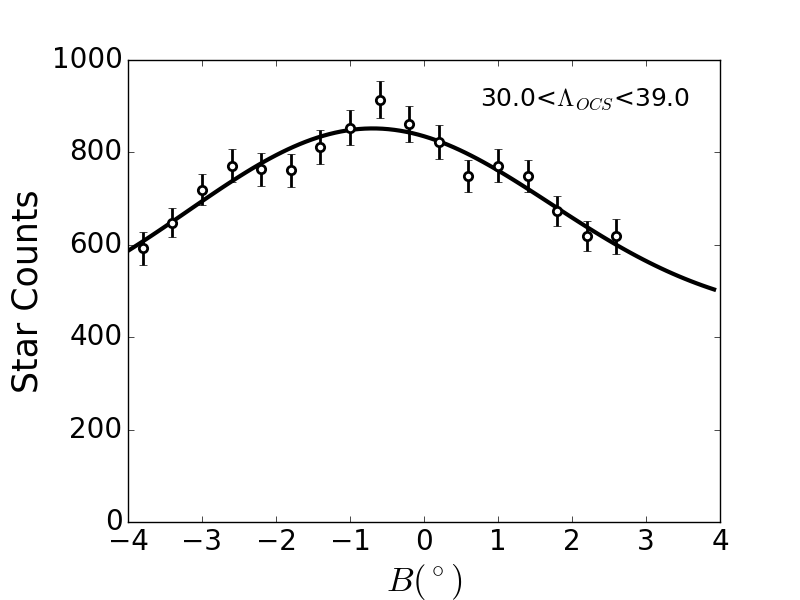}
		\includegraphics[width = 5.9cm]{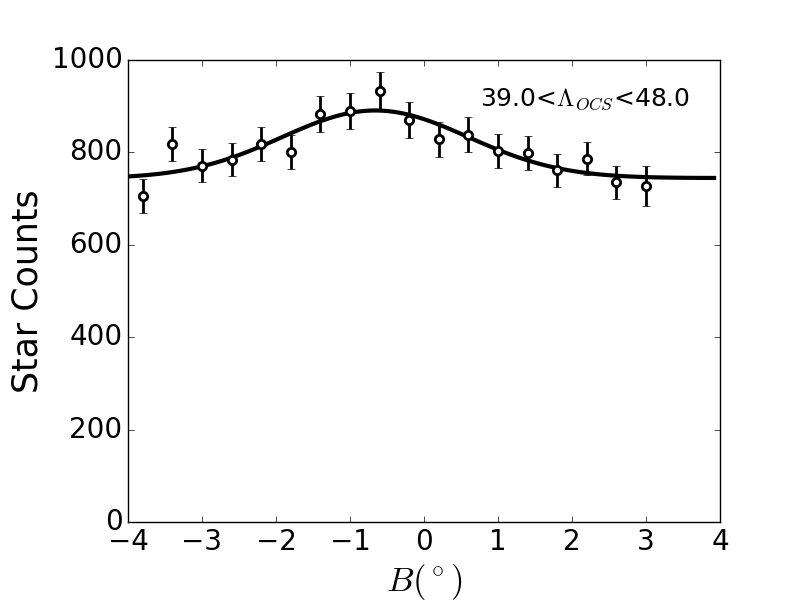}
	    \caption{The distribution of stars within each set of $\Lambda$ bins as a function of $B$. The open circles represent the number of stars in each $B$ sub-bin, and the black line is the curve of best fit using Equation \ref{eq:beta_model}. We excluded $B$ sub-bins from the fit if that sub-bin was less than half filled due to the incomplete areal coverage of the DEC data.}
	\end{figure}\label{fig:beta_disp_fits}
    \end{center}
\begin{center}
    \begin{table}[!ht]
        \centering
        \begin{tabular}{ccc}
             $\Lambda$ Range & Beta Dispersion & Dispersion Error\\
             \hline
             $[-33.0,-24.0]$ & 1.003 & 0.323\\
             $[-24.0,-15.0]$ & 1.001 & 0.208\\
             $[-15.0,-6.0]$ & 1.076 & 0.218\\
             $[-6.0,3.0]$ & 0.612 & 0.146\\
             $[3.0,12.0]$ & 0.791 & 0.308\\
             $[12.0,21.0]$ & 0.947 & 0.257\\
             $[21.0,30.0]$ & 0.886 & 0.193\\
             $[30.0,39.0]$ & 2.263 & 0.825\\
             $[39.0,48.0]$ & 1.230 & 0.275\\
        \end{tabular}
        \caption{Fitted values for the Beta dispersion and dispersion error in each $\Lambda$ range.}
        \label{tab:beta_disp_fits}
    \end{table}
\end{center}
\section{N-Body Simulations}\label{sec:Nbody}

For a given set of dwarf galaxy parameters, we calculate the likelihood that those parameters produce the observed data histogram, as constructed from the values calculated in Section \ref{sec:data}. This section summarizes the methods and algorithms that are laid out in more detail in \cite{shelton2021}.

To calculate the likelihood of a set of progenitor dwarf galaxy parameters, we place the prescribed simulated dwarf galaxy in a static Milky Way potential and let it evolve through time, interacting with with the gravitational well and its own bodies. We employ multi-threading in our N-body algorithm to speed up our calculations. After running the simulation for a prescribed number of timesteps, we calculate the likelihood that the simulated and observed streams result from the same progenitor.

\subsection{Progenitor Creation}\label{sec:dwarf_model}

We represent our simulated dwarf galaxy as a collection of 40,000 bodies; half of the bodies represent visible baryonic matter, and the other half represent dark matter. Regardless of the baryonic and dark matter masses, we employ the same number of bodies, changing the mass per particle to represent the full mass of a component. We choose 40,000 bodies because it is the minimum number of particles needed to consistently generate the same stream over different random seeds (see Figure \ref{fig:like_100000_bodies}).

    \begin{center}
	\begin{figure}[!ht]
	    \centering
		\includegraphics[width = 9.5cm]{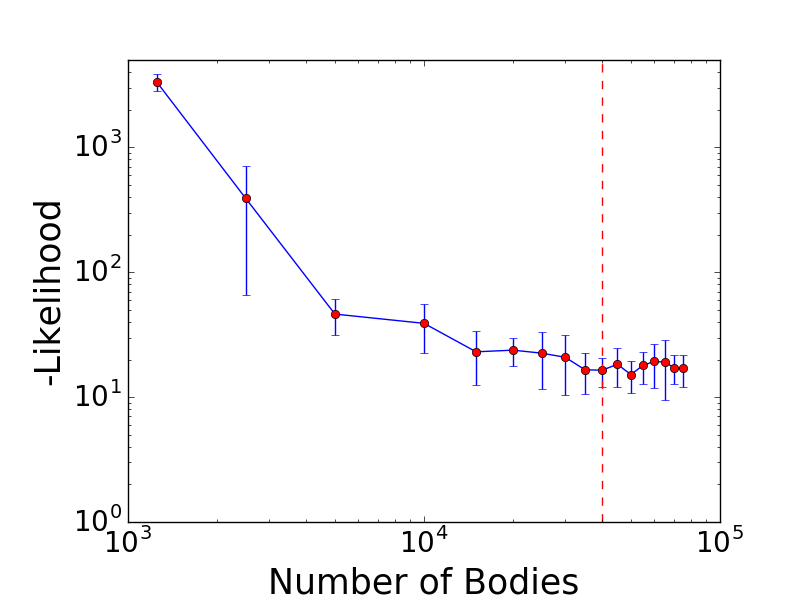}
	    \caption{Plot of average likelihood score as a function of the number of bodies. The likelihood is a measure of how well an individual simulation matches the comparison simulation; zero likelihood is a perfect reproduction. Each point represents the likelihood score of a simulation against a comparison stream generated using the same dwarf parameters averaged over 12 random seeds. Each run calculates the likelihood using the same comparison stream, which was run using 100,000 bodies. The red dashed line is where the number of bodies equals 40,000. Since this is within the area where the likelihood curve flattens, we choose to run 40,000 bodies in all of our simulations.}
	\end{figure}\label{fig:like_100000_bodies}
    \end{center}

To model our progenitor dwarf galaxy, we arrange our 40,000 bodies into two nested, concentric Plummer spheres \citep{plummer_src}. The radial profile of a Plummer sphere is given by:

    \begin{equation}
	\rho(r) = \frac{3M}{4 \pi a^3} \left( 1 + \frac{r^2}{a^2}\right)^{-5/2},
    \end{equation}
where $a$ is a scale radius, and $M$ is the total mass of the Plummer sphere. Each progenitor can thus be described by four parameters, the baryonic mass ($M_B$), the dark matter mass ($M_D$), the baryon scale radius ($a_B$), and the dark matter scale radius ($a_D$). However, since the properties of our dwarf galaxies have a degree of scale invariance, it is more efficient to optimize over the total baryonic component ($M_B$ and $a_B$) and a two ratios ($\xi_R$ and $\xi_M$). These ratios are defined as follows:
    \begin{align}
    	\begin{split}
	        \xi_R &= \frac{a_{B}}{a_B + a_D}\\
	        \xi_M &= \frac{M_{B}}{M_B + M_D}
	    \end{split}.
    \end{align}
The method we use to randomly generate these progenitor dwarf galaxies is detailed in \cite{shelton2021}.

We determine the starting location of the progenitor by first specifying its present-day position and velocity, which we calculated using an orbit for the OCS determined from \cite{newberg2010}. Using its present-day position and velocity, we integrate a single-body orbit backwards in time up to the evolution time $\tau_{evolve}$ to find the progenitor's starting position and velocity, centering the 40,000 body simulated progenitor at that point. Although the orbit remains fixed across simulations, we determined it from different information than we will use to measure the dwarf galaxy parameters; the orbit was fit using sky position, line-of-sight velocity, and heliocentric distance.

Our progenitor dwarf galaxy is thus completely characterized with five parameters: the evolution time $\tau_{evolve}$, the baryonic Plummer radius $a_B$, the radius ratio $\xi_R$, the baryonic Plummer mass $M_B$, and the baryonic mass ratio $\xi_M$. As such, these are the parameters we wish to optimize over using MilkyWay@home.

\subsection{Accelerations on Each Body}

Our Milky Way potential consists of 3 components: a Miyamoto-Nagai disk, a Hernquist bulge, and a logarithmic halo \citep{miyodisk, hern, loghalo}. Our Spherical Hernquist bulge has a scale radius of 0.7 kpc and a mass of = 9.9 $\times$ $10^{10}$ $M_{\odot}$\citep{law2005}. The Miyamoto-Nagai disk in our simulations has a scale radius of 6.5 kpc, a scale height of 0.26 kpc, and a mass of 3.4 $\times$ $10^{10}$ $M_{\odot}$ \citep{law2005}. We implement a logarithmic halo with a circular velocity of 73 km s$^{-1}$ (74.61 kpc Gyr$^{-1}$) \citep{newberg2010} and a scale radius of 12 kpc \citep{law2005} to model the dark matter halo. We translate these potentials into accelerations by calculating the negative gradient of each potential ($\bm{a} = -\nabla\phi)$. Within our simulations, we represent distances in kiloparsecs (kpc), time in gigayears (Gyr), and mass in Structure Masses (SM), where 1 SM  =  222,288.47 $M_{\odot}$. Using these units, $G=1$.

Our algorithm uses classical Newtonian gravitation to calculate the acceleration between two bodies. We employ a Barnes-Hut tree algorithm \citep{TreeCode} which has a complexity of $O(NlogN)$ to speed up calculations. Our N-body simulations employ a Velocity Verlet method \citep{verlet} to calculate the new positions and velocities of each particle at the end of each timestep. We use this Velocity Verlet algorithm because it is a symplectic integrator, meaning the area of its phase space is conserved as the system evolves through time. This property is especially attractive for our integrator since it implies that energy is conserved as the dwarf galaxy falls into the Milky Way, an important constraint that must be maintained throughout our N-body simulations. The calculation of our accelerations, timestep, and softening length is further detailed in \cite{shelton2021}.

\subsection{Phase Space Evolution Method}

After establishing the Milky Way potential and creating the progenitor dwarf, we determine the starting point of our simulation by defining approximately where we expect the progenitor's core to lie in the stream at the present time, and thus where we would like the simulated progenitor core to end up at the end of the simulation. For the OCS, we place this starting point at $(l,b,R)=(258.0^\circ, 45.8^\circ, 21.5  \mbox{ kpc})$ with a velocity  $(v_x, v_y, v_z)=(-185.5, 54.7, 147.4)$ kpc Gyr$^{-1}$. This velocity is given in right-handed Galactic Cartesian coordinates, in which the X-axis is the direction from the Sun to the Galactic Center and the Y-axis is in the direction of the Sun's motion. This starting point defines the orbit of the OCS and was calculated using the orbit fit from \cite{newberg2010}. We place a single body at the starting point and evolve that body backwards in time by our evolution time, $\tau_{evolve}$. We replace the body with the simulated dwarf galaxy progenitor, and the N-body integration begins.

\subsection{Likelihood Calculation}\label{sec:like_calc}

After running a simulation to completion, we need to translate the resulting tidal stream into a histogram to compare with the data histogram. We transform our bodies into a Lambda-Beta ($\Lambda, B$) system that follows the OCS along its $B=0$ great circle. We use the same coordinate system as derived in \cite{newberg2010}. We split the stream into a histogram that consists of 28 $\Lambda$ bins subtending the range ($-33^\circ, 48^\circ$) and one Beta bin covering the range ($-15^\circ, 15^\circ$). Within each bin, we record the number of bodies and normalize the counts with respect to the total number of bodies within the histogram's range. We also calculate the Beta dispersion at each $\Lambda$ where we have a measured stream width.


To compare the similarity between two histograms, we employ a metric called the likelihood score. This value represents the natural logarithm of the probability that two histograms match. As such, the likelihood score ranges from negative infinity (worse case set to -9999999.9) to 0, where a higher likelihood is indicative of a closer match. Our likelihood score consists of 3 components: the Earth Mover Distance (EMD) Component, the Cost Component, and the Beta Dispersion Component.

The EMD Component \citep{emd} measures how well the shapes of two normalized histograms match each other. Given two histograms, one representing the number of stream stars as a function of $\Lambda$ and the other representing the number of simulation bodies as a function of $\Lambda$, the code calculates the minimum amount of ``work" necessary to deform one histogram into the other and translates that work into an EMD score. If the EMD score is greater than some maximum EMD score we define ($EMD_{max}=50$), the code outputs the worst case likelihood score. Otherwise, we calculate the EMD component to be:

\begin{equation}
    \ln(\mathcal{L}_{EMD}) = 300\ln\left(1 - \frac{EMD}{EMD_{max}}\right)
\end{equation}

The Cost Component compares the total mass within each histogram and adds a penalty commensurate to the mass difference. We utilize this component because the EMD score can only be calculated if both histograms are normalized, so the information about their masses lost. The Cost Component is given by:

\begin{equation}
    \ln(\mathcal{L}_{Cost}) = -\frac{\left(M_{sim}N_{sim}-M_{data}N_{data}\right)^2}{2\left(M_{data}^2N_{data} + M_{sim}^2N_{sim}\left(\frac{N_{sim}}{N_{total}}\right)\left(1 - \frac{N_{sim}}{N_{total}}\right)\right)}
\end{equation}
where $M_{sim}$ is the mass per body in the simulation, $M_{data}$ is the mass per turnoff star we calculated in Section \ref{sec:est_mass}, $N_{sim}$ is the number of baryonic bodies within the histogram's range, $N_{data}$ is the number of turnoff stars within the histogram's range (5,631 stars), and $N_{total}$ is the total number of baryonic bodies we ran in the simulation (20,000).


The Beta Dispersion Component compares the width of the streams as a function of $\Lambda$. For each $\Lambda$ bin in the simulation, we calculate the Beta dispersion $\sigma_{\beta,sim,i}$ and its error $\delta_{\sigma_{\beta,sim,i}}$, and then compare them to the values determined from the stellar data, $\sigma_{\beta,data,i}$ and $\delta_{\sigma_{\beta,data,i}}$. The formula for the Beta Dispersion Component is given by:

\begin{equation}
    \ln(\mathcal{L}_{Disp}) = -\frac{1}{2}\sum_i^{N_{bins}}\frac{\left(\sigma_{\beta,sim,i} - \sigma_{\beta,data,i}\right)^2}{\delta_{\sigma_{\beta,sim,i}}^2 + \delta_{\sigma_{\beta,data,i}}^2}.
\end{equation}

The total likelihood score is sum of each of these components:

\begin{equation}
    \ln(\mathcal{L}) = \ln(\mathcal{L}_{EMD}) + \ln(\mathcal{L}_{Cost}) + \ln(\mathcal{L}_{Disp})
\end{equation}
The details of how each component is calculated can be found in \cite{shelton2021}. Only the locations of the baryonic bodies contribute to the calculation of the likelihood score, because we are only able to observe the stars in the tidal stream.

Due to the chaotic behavior of N-body systems, the slightest deviation in the progenitor parameters (or the random seed) can shift the core of our progenitor further ahead or behind in the stream, drastically affecting the likelihood. To mitigate this chaos, we check the likelihood score over a range of evolution times to ``match" the cores of the simulated stream to that of the data. Once the simulation reaches 98\% of the evolution time, the code calculates the likelihood at each timestep and keeps the snapshot with the best likelihood. This comparison continues until the simulation time reaches 102\% of the evolution time. The best likelihood in this range of forward evolution times is returned as the likelihood for the given parameter set.

\section{Optimization Procedure}\label{sec:Optimize}

Exploring our parameter space requires hundreds of thousands of likelihood calculations, each requiring an N-body simulation. Our N-body simulations on average take 15 minutes to compute on a single processor; it would take several years to run this optimization on an average laptop. Therefore, computing these optimized parameters within a reasonable time-frame demands the use of a much more sophisticated computer network.

MilkyWay@home is a collection of roughly 26,000 volunteered computers connected by the Berkeley Open Infrastructure for Network Computing (BOINC), operating at 1.5 PetaFLOPS of combined computing power. MilkyWay@home uses this massive computing power in conjunction with the Time Asynchronous Optimization \citep[TAO;][]{desell2009} package to optimize the parameters which define the size and shape of the progenitor dwarf galaxy. This section explains the differential evolution genetic algorithm we use to optimize our progenitor parameters and describes the parameter space we explore.

\subsection{Differential Evolution Genetic Algorithm}
MilkyWay@home starts an optimization by first generating a random population of 50 parameter sets, each containing the five parameters described in Section \ref{sec:dwarf_model}, on the main server. For each parameter set in the population, MilkyWay@home sends out a package containing an N-body executable, a Lua file containing important settings for the simulation, the aforementioned parameter set, and a comparison histogram containing the real stellar data to several of the 26,000 volunteered computers. The volunteered computer runs the N-body executable with the chosen parameters and generates a simulated tidal stream, calculates the density of bodies across the stream, and logs the information into a histogram. The newly generated histogram is compared to the data histogram, and the computer generates a likelihood score based on how well the two histograms match. This score and its associated parameter set are sent back to the main server.

After calculating the likelihood scores of the initial population, we employ a differential evolution algorithm to generate new parameter sets to test. For each member of our population $\textbf{\textit{x}}$, we select three other random distinct members ($\textbf{\textit{a}}$, $\textbf{\textit{b}}$, and $\textbf{\textit{c}}$) to act as ``genetic donors" to the new parameter set $\textbf{\textit{y}}$. We then generate a random integer $p$ between 1 and $n$, where $n$ is the number of elements in our parameter set (in this case 5). Also, we generate a random number $r_i$ between 0 and 1 for each element $x_i$. Given these numbers, the $i^{th}$ element $y_i$ of the new parameter set is thus defined as:

    \begin{equation}\label{eq:diff_evolution}
	   y_{i} = 
	   \begin{cases}
	   a_{i} + F(b_{i}-c_{i})  \indent &r_i < CR \vee i=p\\
	   x_i &otherwise
	   \end{cases}.
    \end{equation}
where $CR$ is the crossover rate (set to 0.9) and $F$ is the differential weight (set to 0.8). These newly created sets are then sent to our volunteers to calculate a likelihood score. Once a parameter set returns with its likelihood score, it is compared against the rest of the population. Only the parameter sets with the highest likelihood scores are kept to act as ``genetic donors” for new parameter sets.

Ultimately, the population of parameter sets converges to the phase space point with the highest likelihood, and thus, to a set of parameters that most accurately matches the stellar data. We say an optimization has converged when all members of the population are identical. MilkyWay@home can take between two to four weeks to converge.

\subsection{Timestep Cut in Parameter Space}

Although \cite{shelton2021} gives us a realistic physical basis for our choice of timestep, there are still regions of our parameter space which could lead to unwieldy run times. In particular, these occur when our generated dwarf progenitor is extraordinarily dense. MilkyWay@home searches over a 5-dimensional parameter space for the parameters with the highest likelihood (search range defined in Table \ref{table:searchrange}). However, it would not be wise to explore the full 5D box as the corners of our parameter space with high mass and low radius would produce exceedingly dense progenitors which could take up to 80 days to run. Regions of parameter space that would require more than 1.5 million timesteps in the simulation are excluded from our optimization. Figure \ref{fig:phase_space} shows the excised region of parameter space.

    \begin{center}
	\begin{table}[!ht]
	\centering
	    \begin{tabular}{ccccc}
		$\tau_{evolve}$ (Gyrs) & $R_B$ (kpc) & Radius Ratio ($\xi_R$=$\frac{R_B}{R_B + R_D}$) & $M_B$ (SM) & Mass Ratio ($\xi_M$=$\frac{M_B}{M_B + M_D}$)\\
		\hline
		[2.0 - 6.0] & [0.01 - 0.50] & [0.1 - 0.6] & [0.1 - 100.0] & [0.001 - 0.95]\\
	    \end{tabular}
	\caption{The search ranges for our optimizations against the stellar data.}
	\label{table:searchrange}
	\end{table}
    \end{center}

    \begin{center}
	\begin{figure}[!ht]
	    \centering
		\includegraphics[width = 9.5cm]{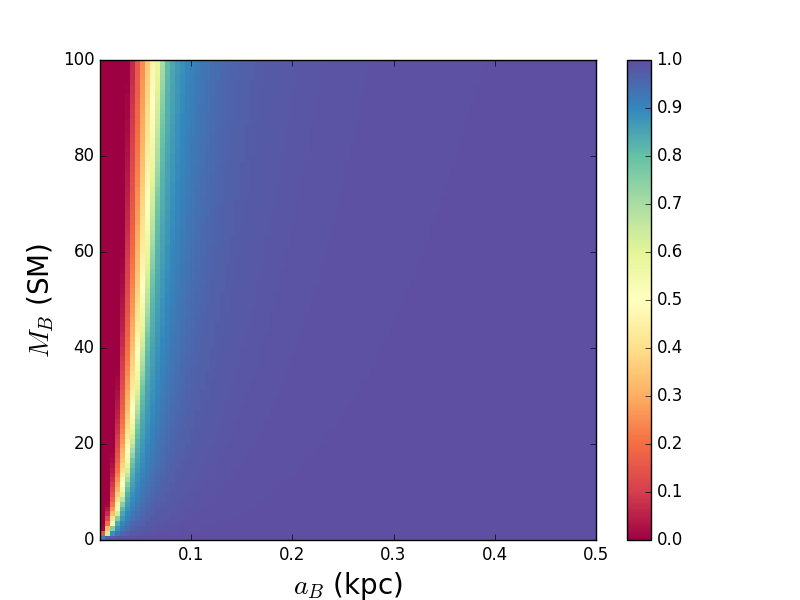}
	    \caption{Plots illustrating the fraction of our 5D parameter space searched as a function of the baryonic Plummer scale radius ($a_B$) and the baryonic progenitor mass ($M_B$). The color of a pixel shows the fraction of the phase space tested for a given ($a_B$,$M_B$). Redder regions in these plots are where a larger percentage of our parameter space is removed. Note that areas where the density of the progenitor galaxy is higher are where we lose most of our phase space. Of the total phase space, 8.92\% of our parameter space is cut out.}
	\end{figure}\label{fig:phase_space}
    \end{center}

Placing a non-linear cut on our phase space like this does remove the ability for MilkyWay@home to fit the densest ultra-compact dwarf (UCD) galaxies. For example, simulating a dwarf galaxy like M60-UCD1, one of the densest dwarf galaxies ever measured \citep{denseDwarf}, would require roughly 17 million timesteps to evolve the progenitor through our gravitational well for 6.0 gigayears. Fortunately, our fits to the OCS do not suggest a UCD progenitor, but rather a dwarf galaxy with a similar radius but over 100 times less mass.

\section{Results}\label{sec:results}

We ran six optimizations over our five dwarf parameters using MilkyWay@home's distributed supercomputer. The raw converged results for each optimization run can be found in Table \ref{tab:params}.
\begin{center}
    \begin{table}[!ht]
        \centering
        \begin{tabular}{ccccccc}
             Run& $\tau_{evolve}$ (Gyrs) & $R_B$ (kpc) & $\xi_R$ & $M_B$ (SM) & $\xi_M$ & $\ln(\mathcal{L})$ \\
             \hline
             1 & 3.6354$\pm$0.0004 & 0.2560$\pm$0.0010 & 0.3455$\pm$0.0014 & 1.017$\pm$0.011 & 0.0413$\pm$0.0005 & -9.604550\\
             2 & 3.6345$\pm$0.0004 & 0.2327$\pm$0.0005 & 0.2543$\pm$0.0010 & 1.146$\pm$0.007 & 0.0179$\pm$0.0003 & -8.809277\\
             3 & 3.6333$\pm$0.0004 & 0.1812$\pm$0.0007 & 0.1828$\pm$0.0009 & 1.223$\pm$0.020 & 0.0126$\pm$0.0003 & -7.726065\\
             4 & 5.3921$\pm$0.0004 & 0.2060$\pm$0.0010 & 0.1315$\pm$0.0015 & 1.635$\pm$0.017 & 0.1445$\pm$0.0006 & -18.033154\\
             5 & 3.6334$\pm$0.0003 & 0.1842$\pm$0.0007 & 0.1820$\pm$0.0010 & 1.251$\pm$0.014 & 0.0119$\pm$0.0002 & -7.997704\\
             6 & 5.3956$\pm$0.0004 & 0.1987$\pm$0.0004 & 0.1305$\pm$0.0021 & 1.682$\pm$0.009 & 0.1507$\pm$0.0016 & -18.982624\\
        \end{tabular}
        \caption{Fitted parameters from MilkyWay@home optimizations. Errors were calculated using Hessian method as described in Section \ref{sec:est_mass}. Likelihoods closer to 0 imply a better fit. Note that the primary source of error is from the navigation of the likelihood surface as measured by the differences between individual runs, and not the statistical error calculated from the Hessian.}
        \label{tab:params}
    \end{table}
\end{center}

From these results, we see that the differential evolution genetic algorithm we used did not consistently converge to the same parameter set. We find that for Runs 4 and 6, the algorithm reached a local maximum whose parameters suggest a lower dark matter concentration. Runs 3 and 5 have the highest likelihood scores and seem to reflect the true maximum in the likelihood surface, whereas Runs 1 and 2 got close to this solution but got failed to optimize further.

Table \ref{tab:other_params} gives the same optimized dwarf parameters as in Table \ref{tab:params}, but measured in physical units. We also directly compare our data histogram against the optimized run with the best likelihood in Table \ref{tab:compare}.

\begin{center}
    \begin{table}[!ht]
        \centering
        \begin{tabular}{cccccc}
             Run & $\tau_{evolve}$ (Gyrs) & $R_B$ (kpc) & $R_D$ (kpc)& $M_B$ ($M_{\odot}$) & $M_D$ ($M_{\odot}$)\\
             \hline
             1 & 3.6354$\pm$0.0004 & 0.2560$\pm$0.0010 & 0.485$\pm$0.004 & (2.26$\pm$0.02)$\times10^{5}$ & (5.24$\pm$0.09)$\times10^{6}$ \\
             2 & 3.6345$\pm$0.0004 & 0.2327$\pm$0.0005 & 0.682$\pm$0.004 & (2.55$\pm$0.02)$\times10^{5}$ & (1.39$\pm$0.02)$\times10^{7}$ \\
             3 & 3.6333$\pm$0.0004 & 0.1812$\pm$0.0007 & 0.810$\pm$0.006 & (2.72$\pm$0.04)$\times10^{5}$ & (2.13$\pm$0.05)$\times10^{7}$ \\
             4 & 5.3921$\pm$0.0004 & 0.2060$\pm$0.0010 & 1.361$\pm$0.019 & (3.63$\pm$0.04)$\times10^{5}$ & (2.15$\pm$0.02)$\times10^{6}$ \\
             5 & 3.6334$\pm$0.0003 & 0.1842$\pm$0.0007 & 0.828$\pm$0.006 & (2.78$\pm$0.03)$\times10^{5}$ & (2.30$\pm$0.05)$\times10^{7}$ \\
             6 & 5.3956$\pm$0.0004 & 0.1987$\pm$0.0004 & 1.323$\pm$0.024 & (3.74$\pm$0.02)$\times10^{5}$ & (2.11$\pm$0.03)$\times10^{6}$ \\
        \end{tabular}
        \caption{Radial profile and mass calculated from fitted parameters. The primary source of error is from the navigation of the likelihood surface as measured by the differences between individual runs, and not the statistical error calculated from the Hessian.}
        \label{tab:other_params}
    \end{table}
\end{center}
\begin{center}
    \begin{table}[!ht]
        \centering
        \begin{tabular}{c|cccc|cccc}
             & \multicolumn{4}{c}{DATA} & \multicolumn{4}{|c}{SIMULATION (RUN 3)}\\ \cline{2-9}
             $\Lambda$ & Norm. Counts & Error & B Disp. & Disp. Error & Norm. Counts & Error & B Disp. & Disp. Error\\
             \hline
             -31.5 & 0.0316 & 0.0072 & ... & ... & 0.0264 & 0.0022 & 1.543 & 0.130\\
             -28.5 & 0.0304 & 0.0078 & 1.003 & 0.323 & 0.0261 & 0.0021 & 1.220 & 0.104\\
             -25.5 & 0.0186 & 0.0081 & ... & ... & 0.0284 & 0.0022 & 1.315 & 0.106\\
             -22.5 & 0.0114 & 0.0089 & ... & ... & 0.0301 & 0.0023 & 1.184 & 0.092\\
             -19.5 & 0.0266 & 0.0091 & 1.001 & 0.208 & 0.0273 & 0.0022 & 1.138 & 0.094\\
             -16.5 & 0.0261 & 0.0095 & ... & ... & 0.0278 & 0.0022 & 1.068 & 0.087\\
             -13.5 & 0.0433 & 0.0101 & ... & ... & 0.0300 & 0.0023 & 1.071 & 0.084\\
             -10.5 & 0.0382 & 0.0105 & 1.076 & 0.218 & 0.0284 & 0.0022 & 0.994 & 0.080\\
             -7.5 & 0.0172 & 0.0109 & ... & ... & 0.0216 & 0.0020 & 0.969 & 0.089\\
             -4.5 & 0.0201 & 0.0117 & ... & ... & 0.0216 & 0.0020 & 0.924 & 0.085\\
             -1.5 & 0.0224 & 0.0127 & 0.612 & 0.146 & 0.0223 & 0.0020 & 0.716 & 0.066\\
             1.5 & 0.0302 & 0.0127 & ... & ... & 0.0207 & 0.0019 & 0.829 & 0.077\\
             4.5 & 0.0227 & 0.0118 & ... & ... & 0.0190 & 0.0018 & 0.764 & 0.075\\
             7.5 & 0.0297 & 0.0108 & 0.791 & 0.308 & 0.0209 & 0.0019 & 0.656 & 0.062\\
             10.5 & 0.0011 & 0.0110 & ... & ... & 0.0190 & 0.0018 & 0.707 & 0.070\\
             13.5 & 0.0261 & 0.0108 & ... & ... & 0.0220 & 0.0020 & 0.502 & 0.046\\
             16.5 & 0.0277 & 0.0109 & 0.947 & 0.257 & 0.0262 & 0.0022 & 0.702 & 0.058\\
             19.5 & 0.0262 & 0.0113 & ... & ... & 0.0287 & 0.0023 & 0.666 & 0.053\\
             22.5 & 0.0325 & 0.0139 & ... & ... & 0.0388 & 0.0026 & 0.610 & 0.038\\
             25.5 & 0.0655 & 0.0146 & 0.869 & 0.195 & 0.0546 & 0.0031 & 0.660 & 0.038\\
             28.5 & 0.0618 & 0.0140 & ... & ... & 0.0679 & 0.0035 & 0.561 & 0.029\\
             31.5 & 0.0586 & 0.0141 & ... & ... & 0.0745 & 0.0036 & 0.710 & 0.035\\
             34.5 & 0.0925 & 0.0145 & 2.606 & 1.217 & 0.0784 & 0.0037 & 0.771 & 0.037\\
             37.5 & 0.0870 & 0.0149 & ... & ... & 0.0800 & 0.0038 & 0.784 & 0.037\\
             40.5 & 0.0655 & 0.0144 & ... & ... & 0.0598 & 0.0033 & 0.887 & 0.049\\
             43.5 & 0.0240 & 0.0152 & 1.325 & 0.311 & 0.0527 & 0.0031 & 0.863 & 0.051\\
             46.5 & 0.0629 & 0.0150 & ... & ... & 0.0468 & 0.0029 & 0.660 & 0.042\\
        \end{tabular}
        \caption{Histograms of stellar density and Beta dispersion along the stream. Left histogram generated from data. Right histogram shows best simulation fit. The number of counts in each bin is normalized for the purposes of the Earth Mover Distance calculation in our likelihood score.}
        \label{tab:compare}
    \end{table}
\end{center}

\subsection{Using Marginalized Parameter Sweeps to Estimate Error}
We use the Hessian method as described in Section \ref{sec:est_mass} to calculate the errors in each of our fitted parameters. However, to accurately calculate these errors, we need our step size to already be comparable to the errors we wish to calculate. To accomplish this, we perform a marginalized 1-dimensional parameter sweep along each dwarf parameter to visualize an adequate step size. In this section, we describe the methods by which we quickly and efficiently calculate marginalized likelihood scores for these parameter sweeps.

Calculating a marginalized likelihood score from N-body simulations is difficult for two reasons. First, the marginalization computation must be completed within the likelihood space, rather than the log(likelihood) space we represent in our version of the likelihood score. This requires us to numerically compute integrals over regions of our parameter space where the likelihood is several orders of magnitude lower than machine precision. Second, any marginalization method we implement over our likelihood space is inherently computationally expensive, requiring us to calculate a complicated 4-dimensional integral over a rough likelihood surface. Calculating one likelihood score from a set of parameters requires an N-body simulation. If we were to use a crude 3-point Gaussian quadrature method to calculate each integral, that would require us to run 81 N-body simulations to calculate the marginalized likelihood for one point of our parameter sweep. Assuming we use about 30 points for our parameter sweep, that brings the total number of simulations to 2,430, requiring one to three months for each parameter sweep.

To address the first issue, we define a special convention for adding together likelihood scores. Given two likelihood scores $\ln(\mathcal{L}_>)$ and $\ln(\mathcal{L}_<)$ (where $\mathcal{L}_> > \mathcal{L}_<$), we define their sum $\ln(\mathcal{L}_{sum})$ to follow the following equation:

\begin{equation}
    e^{\ln(\mathcal{L}_{sum})} = e^{\ln(\mathcal{L}_{>})} + e^{\ln(\mathcal{L}_{<})}.
\end{equation}
Doing some simple math, we can rewrite this equation as:

\begin{equation}
    \ln(\mathcal{L}_{sum}) = \ln(\mathcal{L}_{>}) + \ln\left(1 + e^{\ln(\mathcal{L}_{<})-\ln(\mathcal{L}_{>})}\right)
\end{equation}
Overall, larger likelihood scores dominate the smaller ones. In a double precision computation, if the two likelihood scores differ by more than 64, the smaller likelihood score is completely ignored in the calculation. Therefore, given a list of likelihoods to add, it is most prudent to add together the smaller likelihood scores first.

To calculate the marginalized likelihood score $\ln(\widetilde{\mathcal{L}})$, we use a Monte Carlo integration method using $N$ random points in our parameter space of volume $V$. To determine the minimum number of random points we need for our calculation, we need to know how $N$ affects the error in $\ln(\widetilde{\mathcal{L}})$. Given a list of likelihoods $\{\mathcal{L}\}$, the marginalized likelihood $\widetilde{\mathcal{L}}$ calculated using Monte Carlo is given by the following equation:

\begin{equation}
    \widetilde{\mathcal{L}} = \frac{V}{N}\sum_i \mathcal{L}_i = V\mu_\mathcal{L},
\end{equation}
where $\mu_\mathcal{L}$ is the average of $\mathcal{L}$. From this, it is clear that the main source of error in this equation comes from the calculation of $\mu_\mathcal{L}$, which is simply the error in the mean:

\begin{equation}
    \delta_{\widetilde{\mathcal{L}}} = V\frac{\sigma_\mathcal{L}}{\sqrt{N}},
\end{equation}
where $\sigma_\mathcal{L}$ is the standard deviation of the population that $\mathcal{L}$ was pulled from. Preforming basic error propagation, we find the error in the marginalized likelihood score to be:

\begin{equation}\label{eq:set_err}
    \delta_{\ln(\widetilde{\mathcal{L}})} = \frac{\delta_{\widetilde{\mathcal{L}}}}{\widetilde{\mathcal{L}}} = \frac{\sigma_\mathcal{L}}{\mu_\mathcal{L}\sqrt{N}}.
\end{equation}
It can then be shown that:

\begin{equation}\label{eq:error_marg}
    \delta_{\ln(\widetilde{\mathcal{L}})} = \frac{1}{\sqrt{N-1}}\sqrt{N\frac{\sum \mathcal{L}_i^2}{\left(\sum \mathcal{L}_i\right)^2}-1}.
\end{equation}

Equation \ref{eq:error_marg} is what we use in Figure \ref{fig:param_sweeps} to determine the errors in our parameter sweeps (shaded in blue). However, given this equation, we can generate a rough estimate for what this error looks like on average. We can reorder our list $\{\mathcal{L}\}$ in descending order, where $\mathcal{L}_0$ is the highest likelihood and $\mathcal{L}_1$ is the second highest likelihood. Since we anticipate that each pair of adjacent entries in the reordered list to differs by tens of orders of magnitude, we can make the following approximations:

\begin{equation}
    \sum_i \mathcal{L}_i \approx \mathcal{L}_0 + \mathcal{L}_1 = \mathcal{L}_0\left(1 + k\right), 0<k\leq1,
\end{equation}
\begin{equation}
    \sum_i \mathcal{L}_i^2 \approx \mathcal{L}_0^2 + \mathcal{L}_1^2 = \mathcal{L}_0^2\left(1 + k^2\right), 0<k\leq1,
\end{equation}
where $k=p_1/p_0$. We can make this approximation since we expect only the largest few likelihoods in our list to impact the summations. Substituting these quantities into Equation \ref{eq:error_marg} gives us:

\begin{equation}
    \delta_{\widetilde{\mathcal{L}}} \approx \frac{1}{\sqrt{N-1}}\sqrt{N\frac{\left(1 + k^2\right)}{\left(1 + k\right)^2}-1} = \sqrt{\frac{1 + k^2 - 2k/(N-1)}{\left(1 + k\right)^2}}.
\end{equation}
Taking the limit where $k$ approaches 0, we find that $\delta_{\widetilde{\mathcal{L}}}$ takes the form:

\begin{equation}
    \delta_{\widetilde{\mathcal{L}}} \approx \sqrt{1 - \frac{2k}{N-1}} \approx 1.
\end{equation}
Surprisingly, we find that the effect of $N$ on $\delta_{\widetilde{\mathcal{L}}}$ is incredibly muted in our likelihood space. In fact, looking at the parameter sweeps in Figure \ref{fig:param_sweeps}, we find that the errors associated with each point are extremely close to 1. It is therefore unreasonable to use a bound in $\delta_{\widetilde{\mathcal{L}}}$ to select a minimum value for $N$.

Instead, we should set $N$ to be the minimum number of measurements by which the unbiased value of $\sigma_p$ can be most efficiently estimated to a set percentage. To determine $\sigma_p$ to within 1\%, the unbiased variance $\sigma_p^2$ must be determined to within 2\%. Also, the relationship between $\sigma_p^2$ and the sample variance $s_p^2$ can be estimated using Bessel's correction:

\begin{equation}
    \frac{\sigma_p^2}{s_p^2} = \frac{N}{N-1}.
\end{equation}
Therefore, to know $\sigma_p^2$ to within 2\%, we measure the likelihoods from 50 random points for each Monte Carlo approximation to calculate our parameter sweeps.
    \begin{center}
	\begin{figure}[!ht]
	    \centering
		\includegraphics[width = 5.9cm]{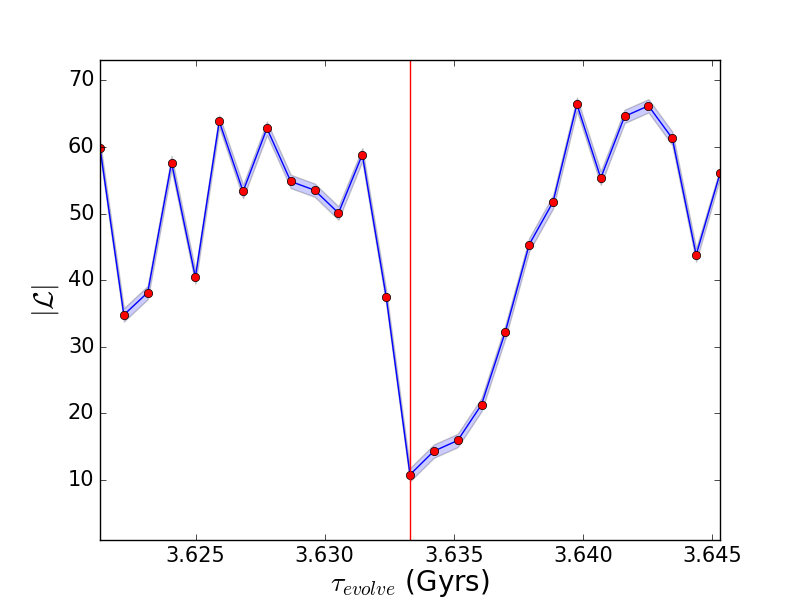}
		\includegraphics[width = 5.9cm]{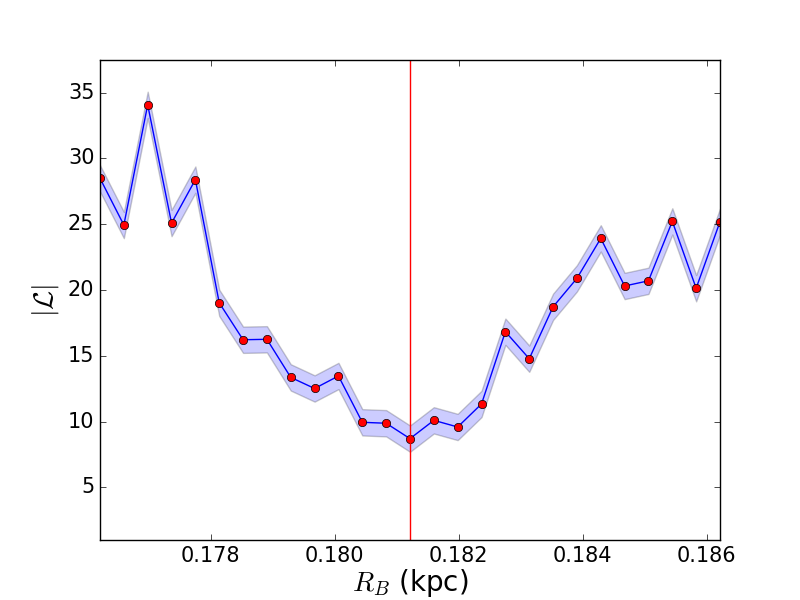}
		\includegraphics[width = 5.9cm]{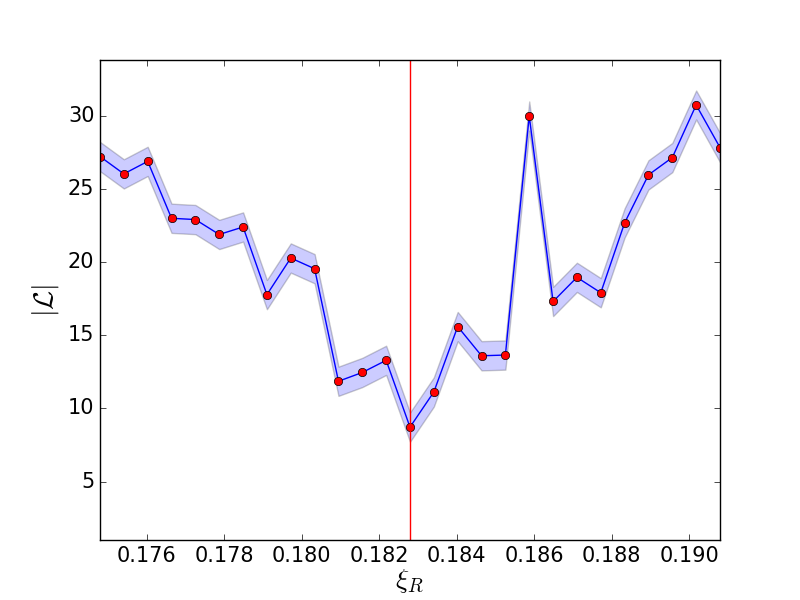}
		\includegraphics[width = 5.9cm]{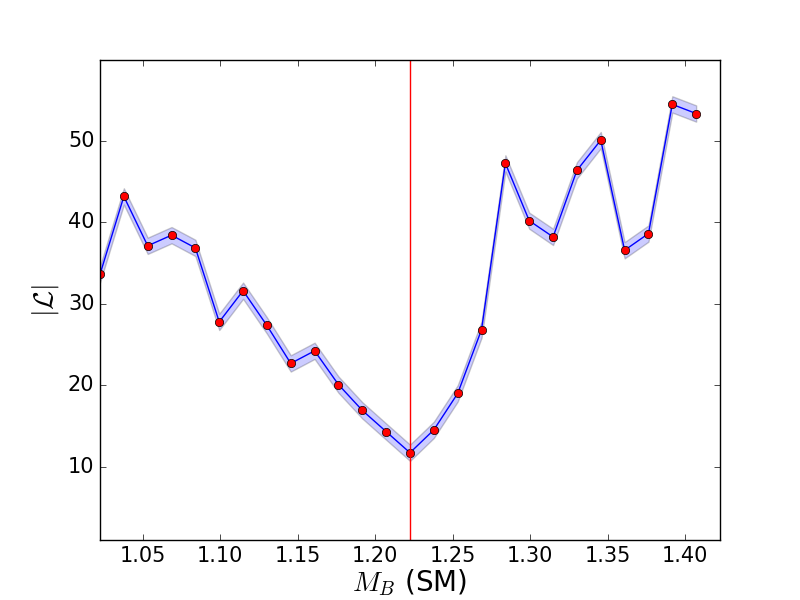}
		\includegraphics[width = 5.9cm]{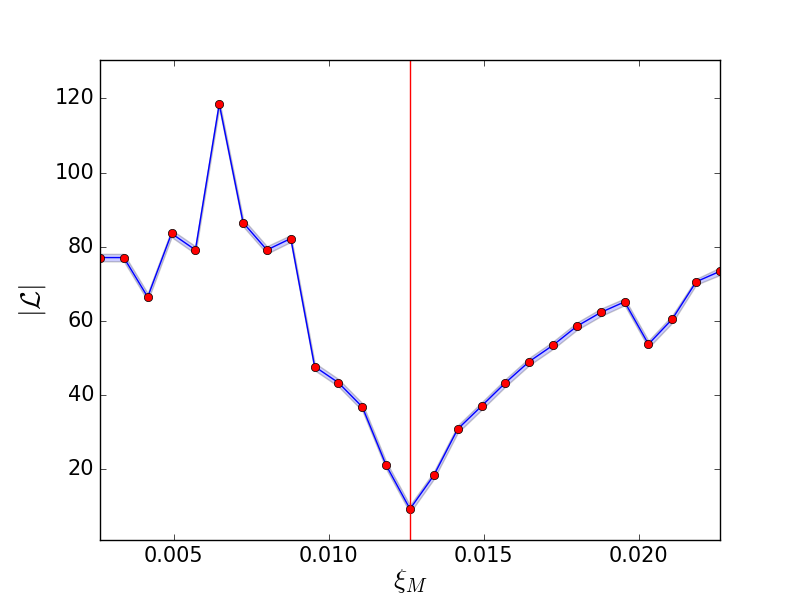}
	    \caption{1-dimensional marginalized parameter sweeps over each dwarf parameter. The blue line shows the negative of the likelihood calculated at each dwarf parameter value while the red line indicates the best fitted value for that parameter (from Run 3). The blue shaded area shows the error in our marginalized likelihood score.}
	\end{figure}\label{fig:param_sweeps}
    \end{center}
    
The likelihood surfaces we observe are not smooth, but we still see an overall dip near the optimal parameters. A Gaussian peak in the likelihood surface translates to a parabola in the log(likelihood) parameter sweeps. Using the apparent width of these peaks, we determine the appropriate step size to use when calculating the Hessian for the purposes of error analysis.

\subsection{Full Result}
After finding the converged values for each of our optimizations, we reran the simulations using each best fit parameter set to check that the generated simulated tidal stream matched the data, as shown in Figures \ref{fig:results} and \ref{fig:results_disp}. In all optimizations, we see a general trend where the $B$ dispersion is relatively well fit for $\Lambda<10^{\circ}$, but poorly fit for $\Lambda>10^{\circ}$, especially near the core. This is likely due to the larger errors in the $B$ dispersion associated with the region near the core in the DEC data at higher $\Lambda_{OCS}$. Most of the optimizations (Runs 2, 3 and 5) converged to similar tidal debris that closely resembles the OCS. Runs 1, 4, and 6 had a poorer likelihood score compared to the other optimizations, and we see that the distribution of bodies near its core is wider across $\Lambda_{OCS}$ than the data. Due to the generally poor fit of these runs, we exclude them from our analysis of the OCS progenitor.

It should be noted that none of these optimizations reproduce the gap ($\sigma=1.06$) in the tidal stream at $\Lambda=10.5^{\circ}$. This is not only because our data can barely resolve this gap, but also because such a gap would only appear as a result of substructures within the Milky Way's dark matter halo \citep{koposov2019}, which is not accounted for in our current halo potential.

Taking the average of the 3 best fitted parameters (Runs 2, 3, and 5), we estimate the following physical properties of the OCS progenitor dwarf galaxy:

\begin{gather*}
        \tau_{evolve} = 3.6337\pm0.0004^{(stdev.)}\pm0.0004^{(Hess.)} \text{Gyrs}\\
        R_B = 0.1994\pm0.0167^{(stdev.)}\pm0.0006^{(Hess.)} \text{kpc}\\
        R_D = 0.773\pm0.046^{(stdev.)}\pm0.005^{(Hess.)} \text{kpc}\\
        M_B = \left(2.68\pm0.07^{(stdev.)}\pm0.03^{(Hess.)}\right) \times 10^5 M_\odot\\
        M_D = \left(1.94\pm0.28^{(stdev.)}\pm0.04^{(Hess.)}\right) \times 10^7 M_\odot\\
        M_{total} = \left(1.97\pm0.28^{(stdev.)}\pm0.04^{(Hess.)}\right) \times 10^7 M_\odot\\
\end{gather*}
where the first error is the standard deviation of the mean and the second error is the propagated error from each individual optimization's Hessian error. We see that our largest source of error comes from the optimization process rather than the curvature of the likelihood surface. So, we only propagate those errors. From these parameters we calculate a fitted mass-to-light ratio of $\gamma=73.5\pm10.6$.

Note that these results and errors are derived under the assumption of a perfect Milky Way potential, orbit, and functional form of the dwarf galaxy and do not include systematic errors due to these assumptions.
    \begin{center}
	\begin{figure}[!ht]
	    \centering
		\includegraphics[width = 5.9cm]{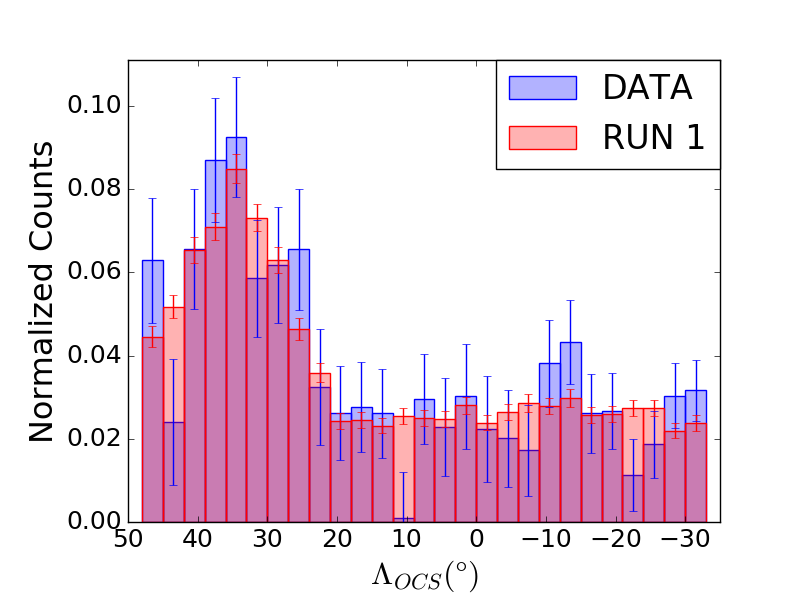}
		\includegraphics[width = 5.9cm]{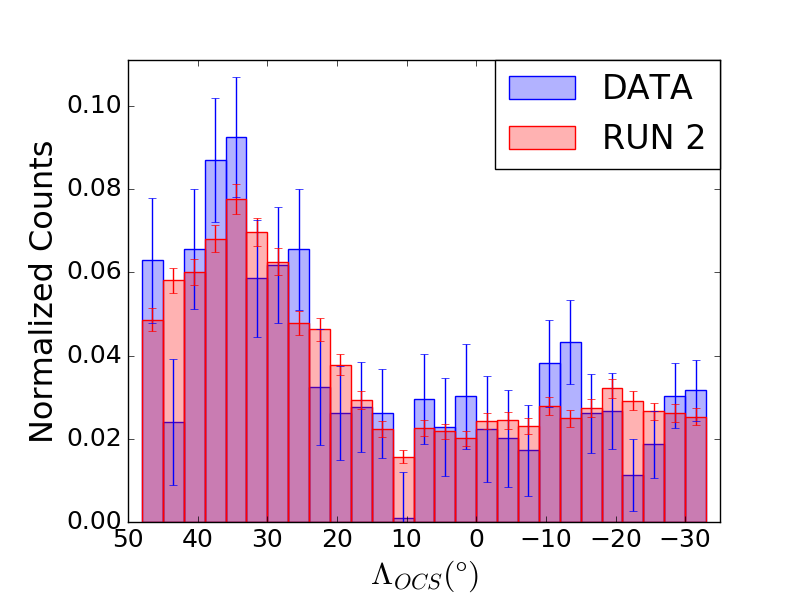}
		\includegraphics[width = 5.9cm]{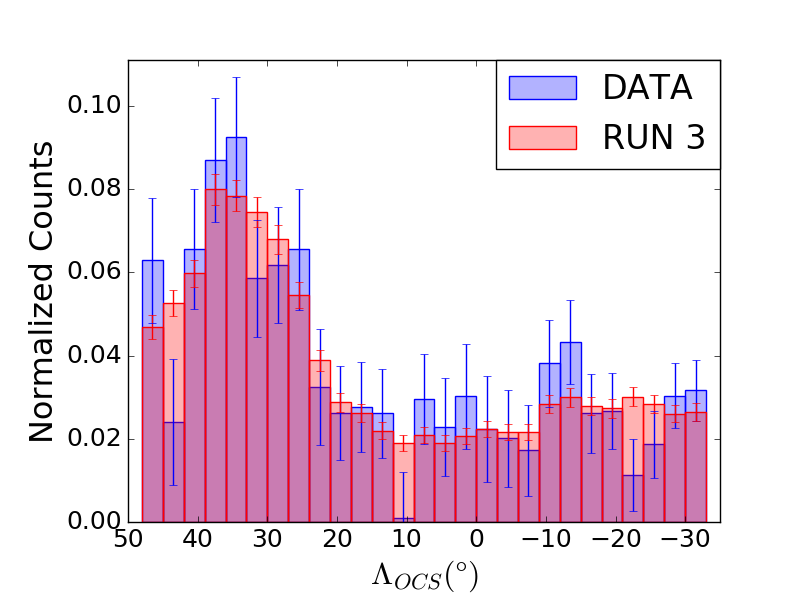}
		\includegraphics[width = 5.9cm]{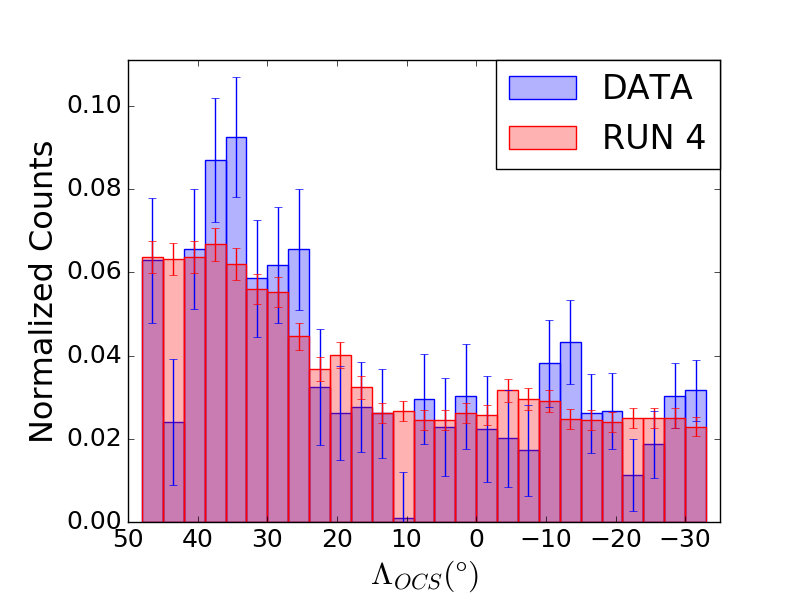}
		\includegraphics[width = 5.9cm]{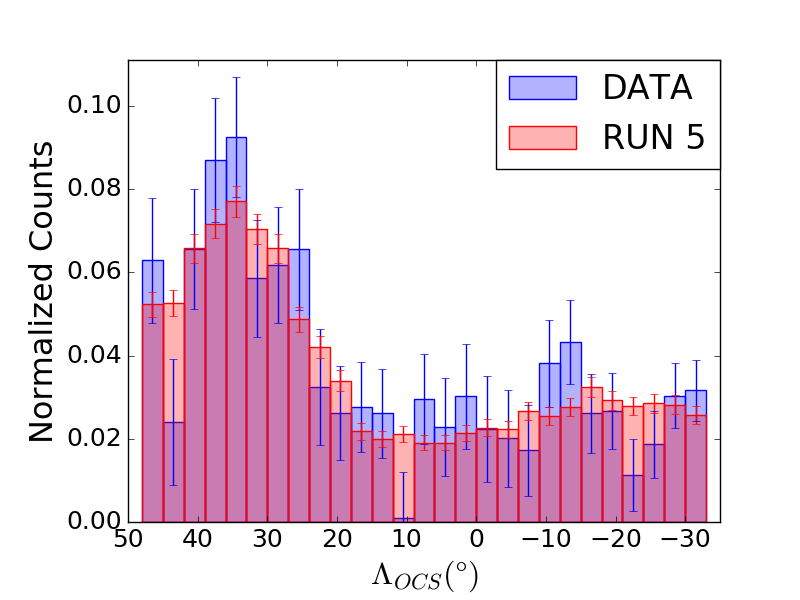}
		\includegraphics[width = 5.9cm]{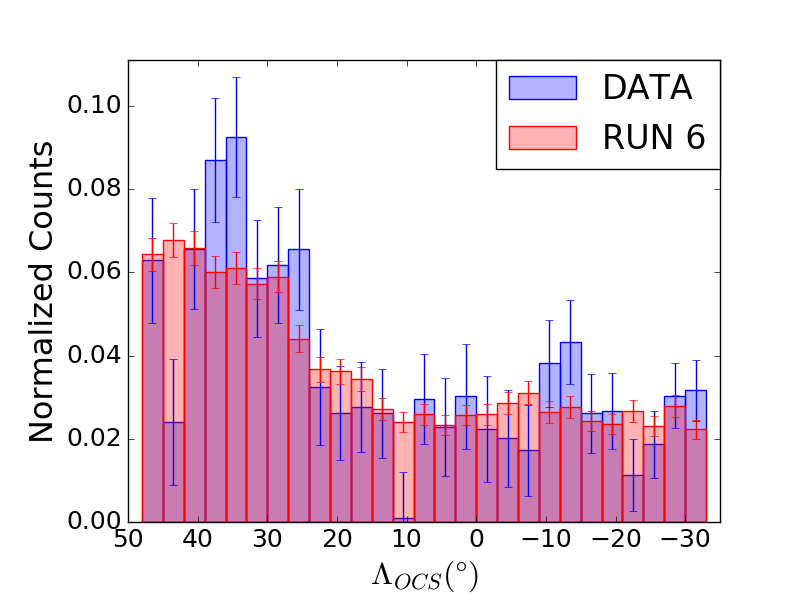}
	    \caption{End state histograms showing normalized body counts over $\Lambda_{OCS}$.}
	\end{figure}\label{fig:results}
    \end{center}
    \begin{center}
	\begin{figure}[!ht]
	    \centering
		\includegraphics[width = 5.9cm]{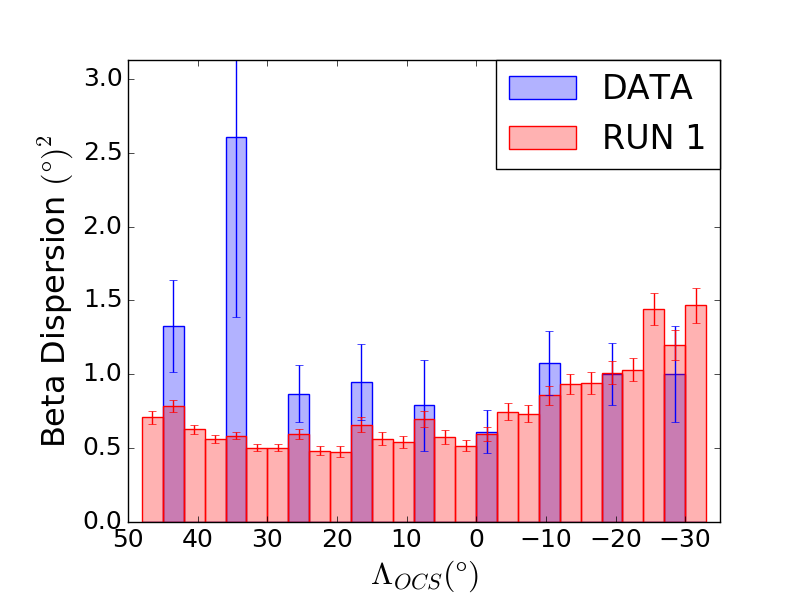}
		\includegraphics[width = 5.9cm]{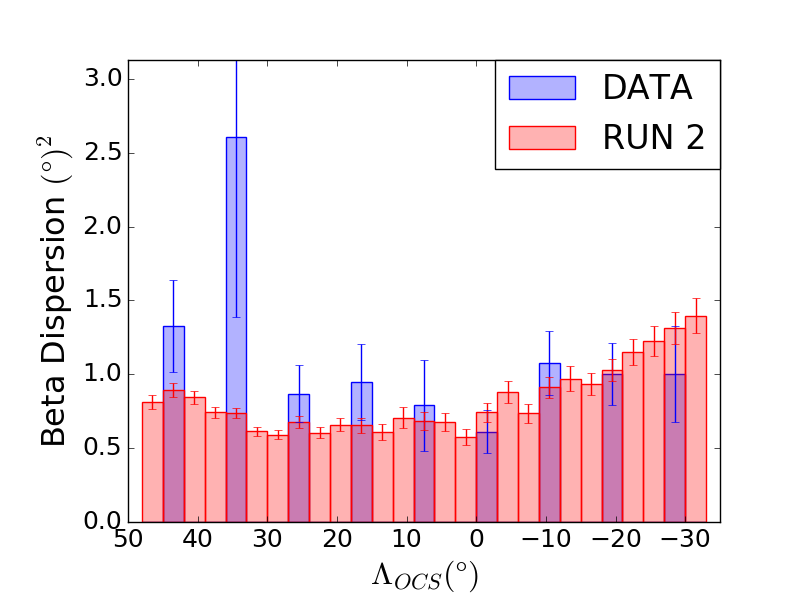}
		\includegraphics[width = 5.9cm]{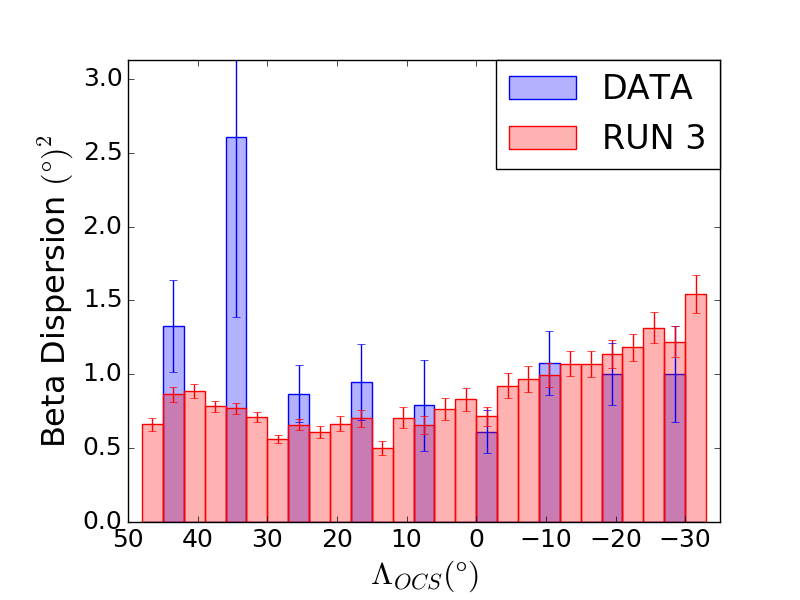}
		\includegraphics[width = 5.9cm]{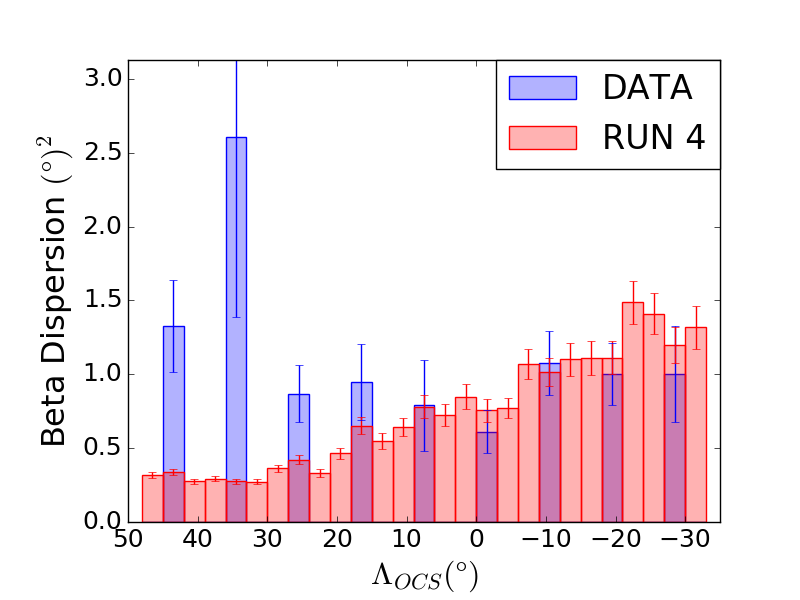}
		\includegraphics[width = 5.9cm]{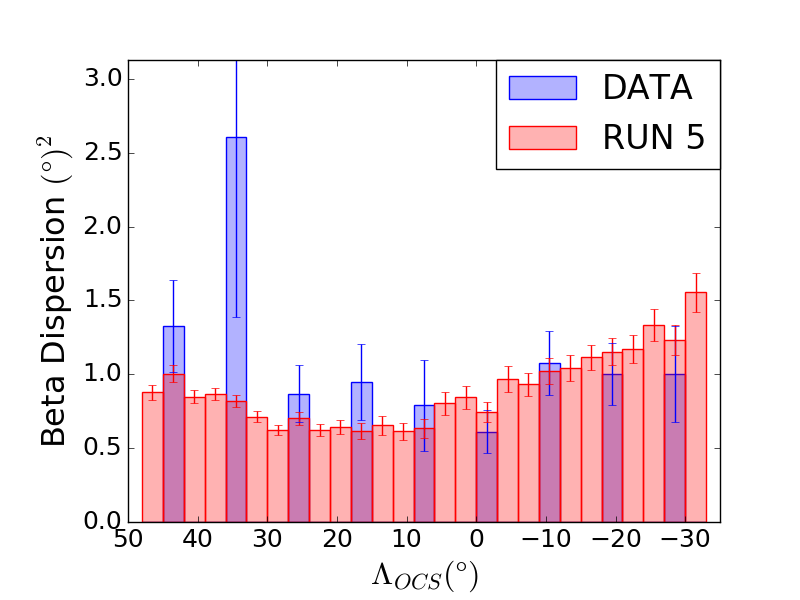}
		\includegraphics[width = 5.9cm]{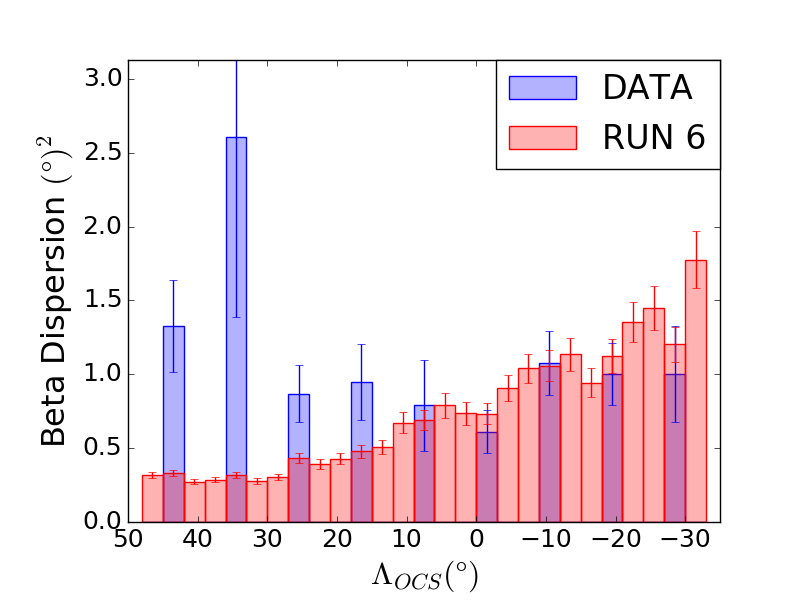}
	    \caption{End state histograms showing $B$ dispersion over $\Lambda_{OCS}$.}
	\end{figure}\label{fig:results_disp}
    \end{center}

\section{Discussion}\label{sec:discuss}
\subsection{An Extremely Low-Mass Diffuse Progenitor}
Our estimated mass for the OCS's progenitor is not consistent with the masses of ultrafaint dwarf galaxies measured from the dispersion of the line-of-sight velocities, which assume spherical symmetry and virial equilibrium. To illustrate this point more clearly, let us use these parameters to calculate the total mass within 300 pc of the progenitor's center. Within this radius, we calculate $(1.6\pm0.1)\times10^5M_{\odot}$ of baryonic matter and $(9.2\pm2.2)\times10^5M_{\odot}$ of dark matter. In total, we find a mass of $(1.1\pm0.2)\times10^6M_{\odot}$ within 300 pc of the progenitor's center, roughly one order of magnitude smaller than what is claimed possible in \cite{strigari2008}. However, these results are only a factor of $\sim2$ smaller than the mass estimated in \cite{newberg2010}, who estimated the total OCS progenitor mass from N-body simulations to be closer to $2.5 \times 10^6 M_{\odot}$. We note that this mass calculation was done under the assumption that the dark matter followed a Plummer profile. However, we note that 300 pc is very close to the half-light radius of the best fit dwarf progenitor, and \cite{shelton2021} showed that our algorithm was most sensitive to the mass within the half-light radius.

\subsection{The Location of the Remnant Progenitor}
We use our fitted simulations to estimate the current location of the OCS's progenitor. Note that there is a peak in the star counts of both the data and the stream simulation around $\Lambda \sim 35^\circ$ in Figure \ref{fig:results}. We know this point of high stellar density must be the location of the progenitor remnant because there is no other reason for stars to pile up at this position, so far from apogalacticon. In this section we will determine the position of the progenitor remnant and show that it is no longer gravitationally bound.

For each optimization, we translate the positions of all the bodies into spherical coordinates ($r,\theta,\phi$), where the origin is the Galactic Center and the z-axis is perpendicular to the Galactic Plane. We bin the baryonic (stellar) bodies in the polar ($\theta$) and azimuthal ($\phi$) angles and select the bin with the highest density of bodies. Using the bodies in the selected bin, we calculate the average Galactocentric distance ($r$) and its standard deviation. To improve the calculation, we remove all bodies with distances more than 2.5 standard deviations away from the mean and recalculate the average, a process known as sigma-clipping. We perform this sigma-clipping 50 times for the sake of being thorough. After calculating the 3D point in space with the highest density of baryonic bodies, we translate it into Galactic and Equatorial coordinates. The progenitor position of each run is shown in Table \ref{tab:progenitor_loc}.
\begin{center}
    \begin{table}[!ht]
        \centering
        \begin{tabular}{ccccc}
             Run & $l$ ($^\circ$) & $b$ ($^\circ$) & $\alpha$ ($^\circ$)& $\delta$ ($^\circ$)\\
             \hline
             2 & 264.0 & 44.4 & 165.7 & -10.1 \\
             3 & 270.4 & 40.5 & 167.7 & -15.9 \\
             5 & 260.4 & 46.0 & 164.6 & -7.3 \\
        \end{tabular}
        \caption{The current sky position of the progenitor in both Galactic and Equatorial coordinates calculated from each optimization.}
        \label{tab:progenitor_loc}
    \end{table}
\end{center}
Our simulations predict that the remnant progenitor core is expected to be at around $(l,b)=((264.9\pm2.9)^\circ,((43.6\pm2.8)^\circ)$, or $(\alpha,\delta)=((166.0\pm0.9)^\circ,(-11.1\pm2.5)^\circ)$. This is close to the sky position where \cite{grillmair2015} detected the OCS's possible missing progenitor, $(\alpha,\delta)\approx(167^\circ,-14^\circ)$.

We next check whether or not the simulated remnant progenitor is still gravitationally bound. For each optimization, we calculate the total kinetic energy of the remnant and compare it to its gravitational potential energy. We select bodies that are within 0.5 kpc of the progenitor's center since that distance is roughly the cylindrical radius of our simulated stream (see Figure \ref{fig:progenitor_zoom}). We transform our bodies into the center-of-momentum reference frame and add together the kinetic energies $(E_K=\frac{1}{2}\sum_i^Nm_i{v_i}^2)$ of all the bodies. Similarly, we add together the gravitational energies from each pair of bodies $(E_P=\sum_i^N\sum_{i<j}^N\frac{Gm_im_j}{|r_i-r_j|})$. We record the energies we calculate for each optimization in Table \ref{tab:energy_bound}.
\begin{center}
    \begin{table}[!ht]
        \centering
        \begin{tabular}{ccc}
             Run & $E_K$ ($SM kpc^{2} Gyr^{-2}$) & $E_P$ ($SM kpc^{2} Gyr^{-2}$)\\
             \hline
             2 & 5.977 & -0.096 \\
             3 & 7.587 & -0.092 \\
             5 & 5.751 & -0.093 \\
        \end{tabular}
        \caption{The kinetic and gravitational potential energies of the simulated remnant progenitor in each optimization. Note that the kinetic energy is much larger than the gravitational potential energy, indicating that the progenitor is no longer gravitationally bound.}
        \label{tab:energy_bound}
    \end{table}
\end{center}
We find that the simulated remnant progenitor has a kinetic energy of (6.4$\pm$0.6) SM kpc$^{2}$ Gyr$^{-2}$ and a gravitational potential energy of -(0.094$\pm$0.001) SM kpc$^{2}$ Gyr$^{-2}$, and is thus gravitationally unbound. Looking at the stream directly in Figure \ref{fig:progenitor_zoom} makes this readily apparent.

    \begin{center}
	\begin{figure}[!ht]
	    \centering
		\includegraphics[width = 8cm]{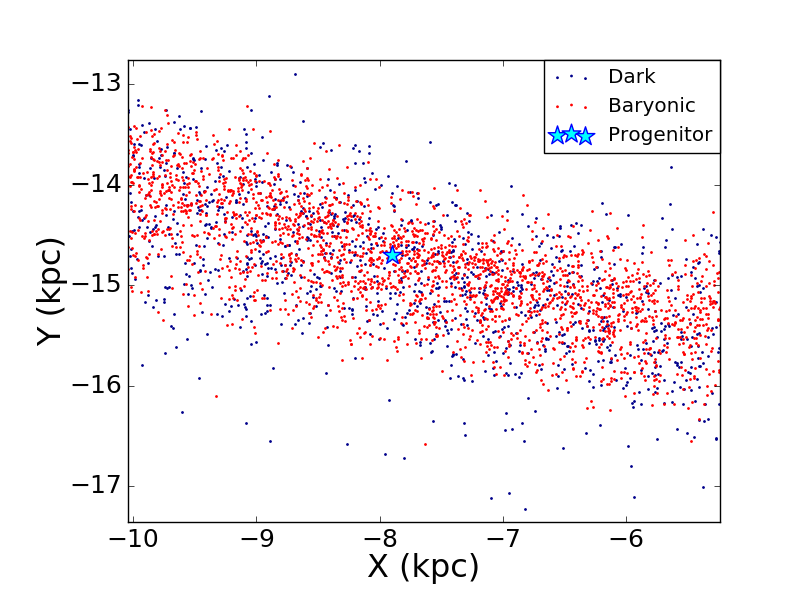}
	    \caption{Close-up view of simulated progenitor remnant (Run 3). The remnant has been fully disrupted and the progenitor's original structure cannot be easily observed.}
	\end{figure}\label{fig:progenitor_zoom}
    \end{center}

\subsection{The Distribution of Dark Matter along the Stream}
The tidal debris we observe in our simulations reveals some interesting aspects of the distribution of baryonic and dark matter in the stellar stream. We analyze the distribution of matter within the tails of the simulated OCS. Figure \ref{fig:nbody_DM} shows that the tails of the OCS are thicker than the central region nearer to the progenitor remnant. There are two possible contributions to this thickening: either the tails are naturally fanning out at apogalacticon due to the symplectic (phase space area preserving) property of gravitational systems, or material with a higher velocity dispersion in the progenitor is being tidally stripped first, populating the regions of the stream further from the core.
    \begin{center}
	\begin{figure}[!ht]
	    \centering
		\includegraphics[width = 5.9cm]{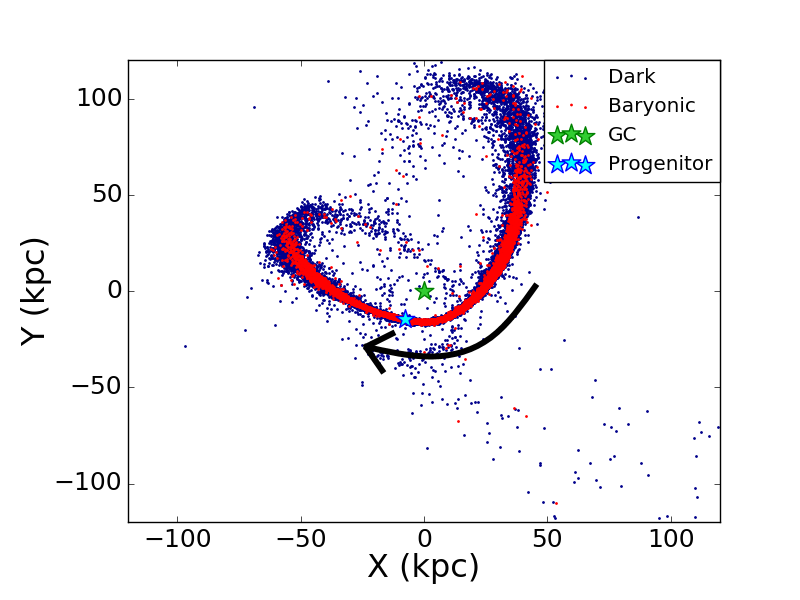}
		\includegraphics[width = 5.9cm]{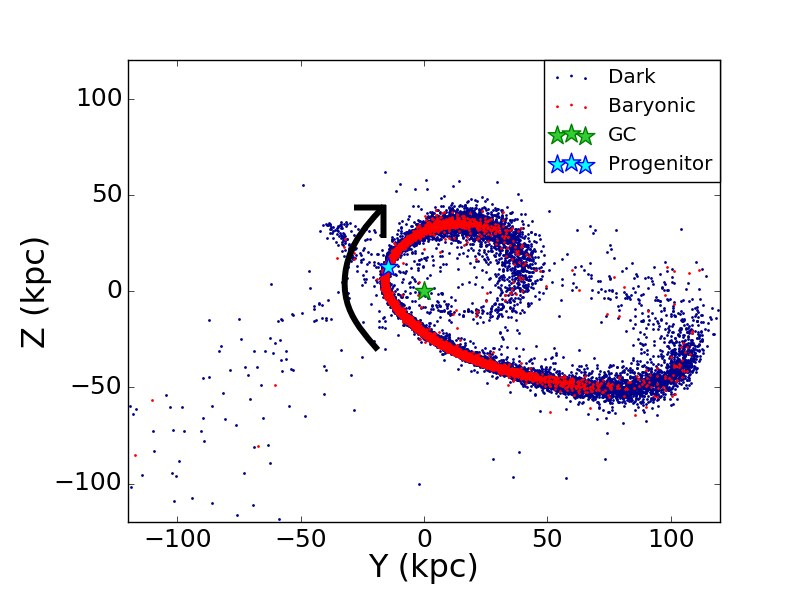}
		\includegraphics[width = 5.9cm]{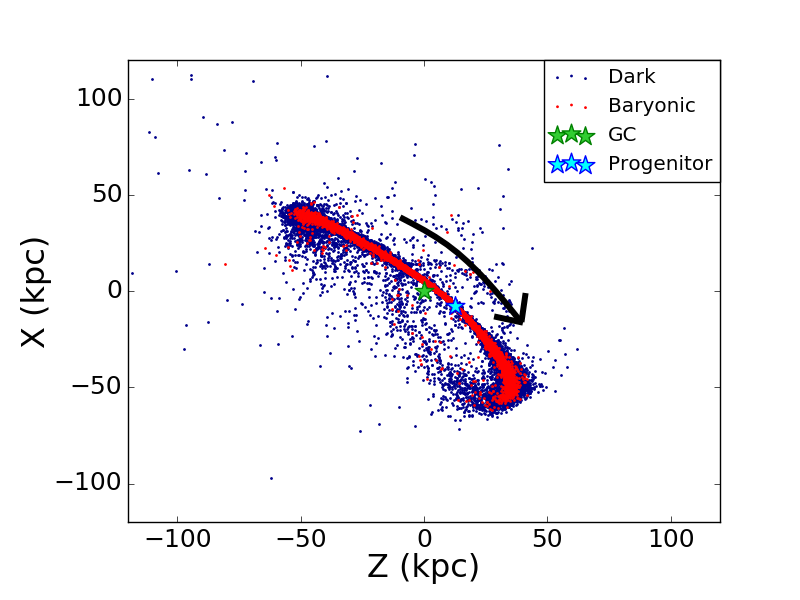}
	    \caption{Tidal stream produced by Run 3 viewed from different angles. We see that the tidal tails of the stream are thicker near apogalacticon. The green star is the Galactic Center (GC) and the cyan star is the location of the OCS's progenitor dwarf galaxy.}
	\end{figure}\label{fig:nbody_DM}
    \end{center}

We can test this by looking at the simulated tidal stream generated using Run3, our best run, when the core is at apogalacticon. If a high velocity dispersion has a stronger impact on the thickness of the tails than the fanning at apogalacticon, then the tails should maintain their thickness even when they are not at apogalacticon. However, as can be clearly seen in Figure \ref{fig:nbody_apo}, while the core is at apogalacticon it bears a similar thickness to the tails in the fully evolved stream. Also, the tails of the stream in the core-apogalacticon state share a similar thickness to the core in the final state. This indicates that the thickening of the tails in the final state is mostly caused by the natural fanning that occurs at apogalacticon.
    \begin{center}
	\begin{figure}[!ht]
	    \centering
	    \includegraphics[width = 8cm]{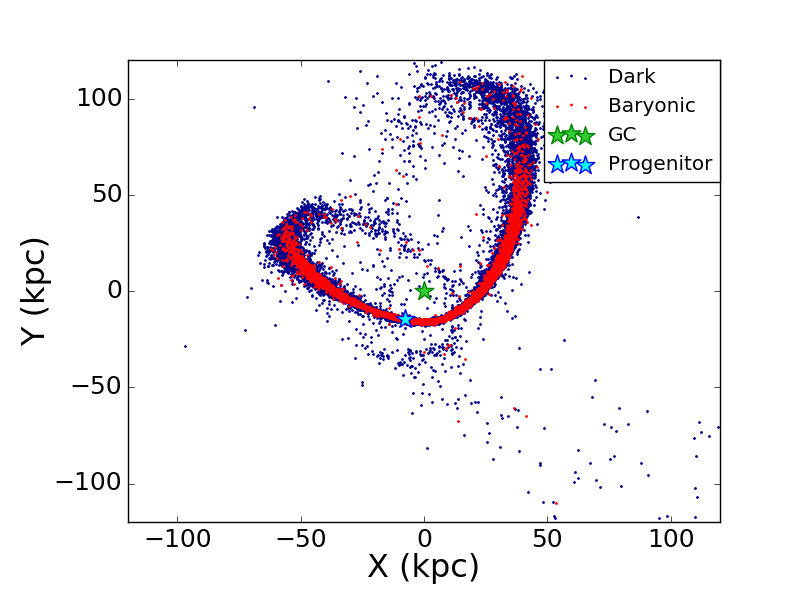}
		\includegraphics[width = 8cm]{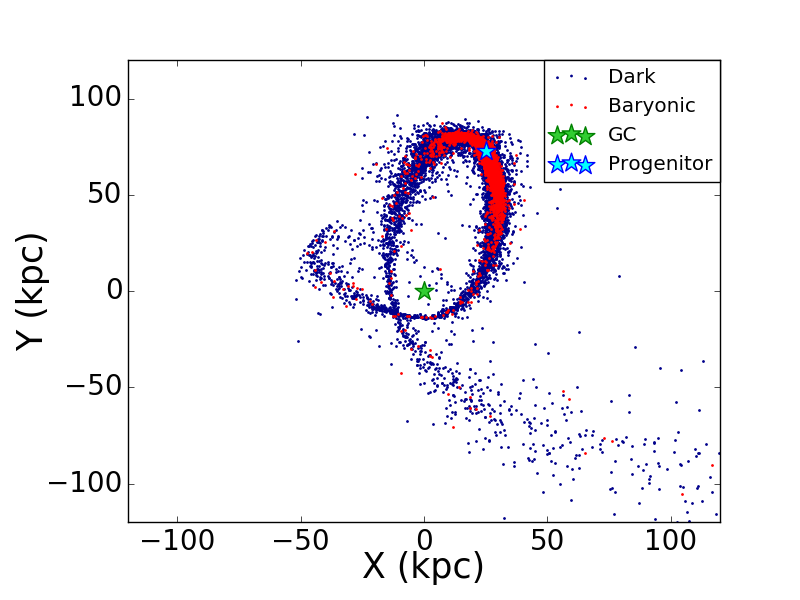}
	    \caption{The final state stream (left) compared to the stream while its progenitor is at apogalacticon (right). The thickness of the stream at apogalacticon is comparable in both images regardless of where the progenitor is.}
	\end{figure}\label{fig:nbody_apo}
    \end{center}

In addition to being thicker than the core, the tails also possess a higher mass-to-light ratio. We see this in Figure \ref{fig:run3_DM}, which maps out the density of baryons and dark matter along $\Lambda_{OCS}$. While the core of the OCS has a relatively high dark matter density, it is still not as high as in its tails, where the mass-to-light ratio jumps to as high as 142 in Run 3. Due to its high dark matter concentration and low baryonic contamination, the tails of the OCS might serve as a better candidate for indirect dark matter detection than the central core.
    \begin{center}
	\begin{figure}[!ht]
	    \centering
		\includegraphics[width = 8cm]{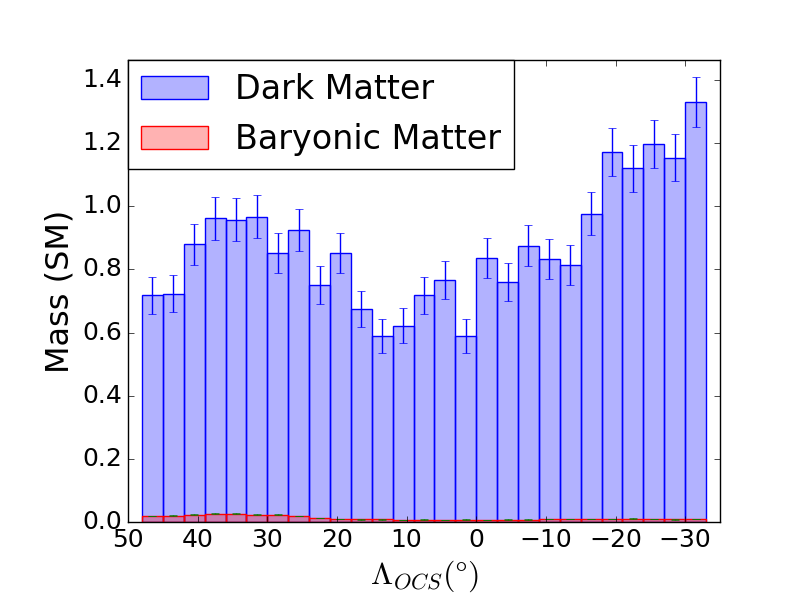}
		\includegraphics[width = 8cm]{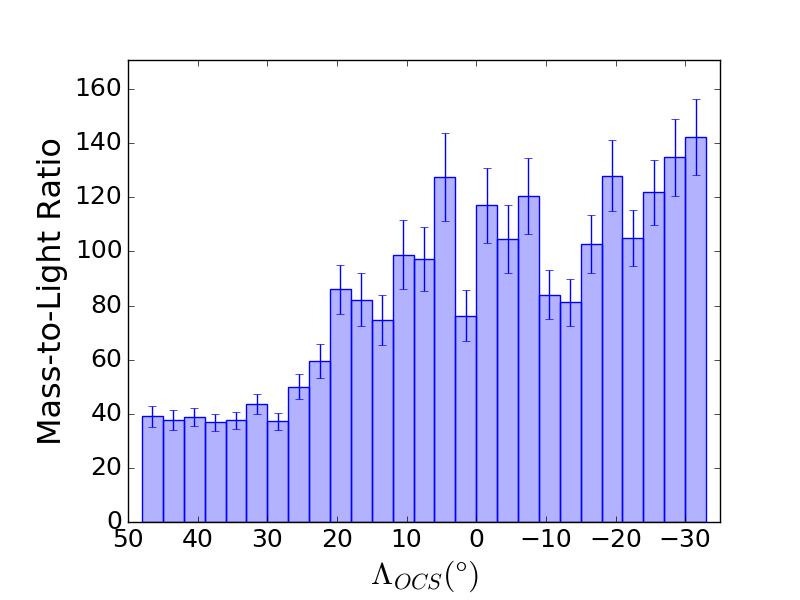}
	    \caption{Distribution of baryons and dark matter in Run 3. The tails clearly have a much higher mass-to-light ratio at points that are farther away from the core.}
	\end{figure}\label{fig:run3_DM}
    \end{center}

\subsection{The Effect of the LMC on the OCS}
When creating our data histogram, we excluded the southern tail of the OCS because our current N-body simulations do not account for the gravitational influence of the LMC. While the OCS's northern tail is not greatly influenced by the presence of the LMC, its southern tail exhibits peculiar perturbations in its stellar velocities which cannot be explained by a static Milky Way potential or a low mass LMC \citep{erkal2019}. While this paper does not fit data from the southern tail, we do plan on including the tail and the gravitational effects of the LMC in a future MilkyWay@home paper.

\subsection{Sources of Systematic Error}
While we measured the OCS progenitor properties without the assumption of dynamical equilibrium, our measurement still relies on a few key presuppositions. First, we assume the Milky Way has a static, axisymmetric Galactic potential for the entirety of the simulation. There currently exists no mechanism in our N-body simulations for our progenitor to perturb the Milky Way to give rise to second-order effects. We do not include any time-dependence in our Milky Way potential. We do not take into account other gravitating systems which could interact with the OCS, such as the LMC or other Milky Way satellites. Adding the LMC to our simulations would not only further perturb the stream, but also induce a reflex motion in the entirety of the Milky Way Galaxy, forcing the dwarf galaxy progenitor to fall into an accelerating potential. It has also been recently discovered that most of the mass in the inner disk of the Milky Way is located in its fast rotating bar \citep{portail2017}, which is also not included in our Galactic model.

Second, we fix the orbit of our simulated OCS throughout each of our optimizations to the values determined in \cite{newberg2010}. Since the orbit is fit to the northern (leading) tail of the OCS, which should be at lower energy than the orbit of the dwarf galaxy itself, we expect the orbit is not exact. We have not explored how the orbit affects the fit properties of the dwarf galaxy progenitor. However, we are only using the stellar distributions along and across the stream, which are not thought to be strongly affected by small differences in the orbit, to constrain the stream. Additionally, the stream created in our best fit simulation is similar to the stream observed, which gives us some confidence that our orbit and progenitor properties are reasonable.

Third, we assume that the OCS progenitor has the form of a two component Plummer sphere. We used a Plummer profile for our progenitor because it is a simple model with an analytic energy distribution function. However, it is probably not the best (and certainly not a perfect) model, particularly for the dark matter halo. We are unsure whether using a different profile, such as a Hernquist or Navarro-Frenk-White profile, would significantly affect the optimization results.

Finally, there is the measurement of baryonic mass in the OCS. As discussed in Section \ref{sec:ngc5053_error}, using the redder NGC 5053 to estimate the mass of the bluer OCS overestimates the mass per turnoff star. From our analysis, we find that the stellar mass could be reduced by a factor of 4 due to inaccuracies in our calculation of the stellar mass per turnoff. This would affect the mass to light ratio, but would not significantly change the total mass within 300 pc of the center of the progenitor.

\section{Conclusion and Future Research}\label{sec:conclusion}
We made the first estimate of the mass and radial profile of the stars and dark matter in the dwarf galaxy progenitor of a tidal stream. To do this, we developed a procedure for characterizing the stellar distribution along and across the OCS, using turnoff stars in the OCS observed in the SDSS and the DEC. We used turnoff stars in NGC 5053 to approximate the stellar mass of the OCS. Using this information for a real observed tidal stream as input to MilkyWay@home, which uses the N-body optimization method developed in \cite{shelton2021}, we calculate the following properties for the dwarf galaxy progenitor of the OCS:

\begin{gather*}
        \tau_{evolve} = (3.6337\pm0.0004) \text{Gyrs}\\
        R_B = (0.20\pm0.02) \text{kpc}\\
        R_D = (0.77\pm0.05) \text{kpc}\\
        M_B = \left(2.68\pm0.07\right) \times 10^5 M_\odot\\
        M_D = \left(1.9\pm0.3\right) \times 10^7 M_\odot\\
        M_{total} = \left(2.0\pm0.3\right) \times 10^7 M_\odot\\
\end{gather*}
From these numbers, we calculated a mass-to-light ratio of $\gamma=73.5\pm10.6$. The implied $(1.1\pm0.2)\times10^6M_{\odot}$ within 300 pc of the progenitor's center is one order of magnitude smaller than the presumed minimum mass of ultrafaint dwarf galaxies.

Our simulations of the OCS's tidal debris show an unbounded and heavily disrupted progenitor remnant at a current sky position of around $(l,b)=((264.9\pm2.9)^\circ,((43.6\pm2.8)^\circ)$, or $(\alpha,\delta)=((166.0\pm0.9)^\circ,(-11.1\pm2.5)^\circ)$. Further studies of the simulated tidal stream suggest that most of the mass of the OCS (especially its dark matter) may reside in its tails, making them optimal candidates for indirect dark matter detection experiments.

While we provide a best fit measurement of the OCS's mass and radial profile, it should be noted that this is only a preliminary fit. There still exist a significant number of variables which need to be addressed before we can definitively measure these stream properties. The most important of these is the gravitational influence of the LMC. \cite{erkal2019} used perturbations in the southern tail of the OCS to place the mass of the LMC at around $1.38^{+0.27}_{-0.24} \times 10^{11} {M_\odot}$, roughly $10\%$ of the Milky Way's mass. It is thus necessary for us to include the gravitational contributions of the LMC in order to properly simulate the infall of the OCS's progenitor.

We also wish to implement a more rigorous method by which to define the orbit of the OCS. In our optimizations, we set a constant orbit using the stellar data from \cite{newberg2010}, assuming the tidal stream itself is representative of the progenitor's orbit. The northern leading tail has a lower energy than the progenitor dwarf galaxy, and thus is expected to trace out a different orbital path. In future work, we plan to fit the orbit simultaneously with the progenitor's mass and radial profile. This will also require us to verify that it is indeed possible to simultaneously fit the five dwarf galaxy parameters and five parameters describing its orbit from fitting simulations to tidal debris.

In future papers, we also plan on evaluating the possible sources of error intrinsic to the N-body simulations themselves as well as the simplifying assumptions we have in this paper. We will explore how different Galactic models representing the Milky Way shift the fitted dwarf parameters and whether different assumed dwarf galaxy radial profiles affect the best fit dark matter mass.

\section{Acknowledgements}

This work was supported by NSF grant AST19-08653; the NASA/NY Space Grant; contributions made by the Marvin Clan, Babette Josephs, and Manit Limlamai; and the 2015 Crowd Funding Campaign to support Milky Way research.

Funding for the SDSS and SDSS-II has been provided by the Alfred P. Sloan Foundation, the Participating Institutions, the National Science Foundation, the U.S. Department of Energy, the National Aeronautics and Space Administration, the Japanese Monbukagakusho, the Max Planck Society, and the Higher Education Funding Council for England. The SDSS Web Site is http://www.sdss.org/.

The SDSS is managed by the Astrophysical Research Consortium for the Participating Institutions. The Participating Institutions are the American Museum of Natural History, Astrophysical Institute Potsdam, University of Basel, University of Cambridge, Case Western Reserve University, University of Chicago, Drexel University, Fermilab, the Institute for Advanced Study, the Japan Participation Group, Johns Hopkins University, the Joint Institute for Nuclear Astrophysics, the Kavli Institute for Particle Astrophysics and Cosmology, the Korean Scientist Group, the Chinese Academy of Sciences (LAMOST), Los Alamos National Laboratory, the Max-Planck-Institute for Astronomy (MPIA), the Max-Planck-Institute for Astrophysics (MPA), New Mexico State University, Ohio State University, University of Pittsburgh, University of Portsmouth, Princeton University, the United States Naval Observatory, and the University of Washington.

Funding for the Sloan Digital Sky Survey IV has been provided by the Alfred P. Sloan Foundation, the U.S. Department of Energy Office of Science, and the Participating Institutions.

SDSS-IV acknowledges support and resources from the Center for High Performance Computing  at the University of Utah. The SDSS website is www.sdss.org.

SDSS-IV is managed by the Astrophysical Research Consortium for the Participating Institutions of the SDSS Collaboration including the Brazilian Participation Group, the Carnegie Institution for Science, Carnegie Mellon University, Center for Astrophysics | Harvard \& Smithsonian, the Chilean Participation Group, the French Participation Group, Instituto de Astrof\'isica de Canarias, The Johns Hopkins University, Kavli Institute for the Physics and Mathematics of the Universe (IPMU) / University of Tokyo, the Korean Participation Group, Lawrence Berkeley National Laboratory, Leibniz Institut f\"ur Astrophysik Potsdam (AIP),  Max-Planck-Institut f\"ur Astronomie (MPIA Heidelberg), Max-Planck-Institut f\"ur Astrophysik (MPA Garching), Max-Planck-Institut f\"ur Extraterrestrische Physik (MPE), National Astronomical Observatories of China, New Mexico State University, New York University, University of Notre Dame, Observat\'ario Nacional / MCTI, The Ohio State University, Pennsylvania State University, Shanghai Astronomical Observatory, United Kingdom Participation Group, Universidad Nacional Aut\'onoma de M\'exico, University of Arizona, University of Colorado Boulder, University of Oxford, University of Portsmouth, University of Utah, University of Virginia, University of Washington, University of Wisconsin, Vanderbilt University, and Yale University.

This project used data obtained with the Dark Energy Camera (DECam), which was constructed by the Dark Energy Survey (DES) collaboration. Funding for the DES Projects has been provided by the DOE and NSF(USA), MISE(Spain), STFC(UK), HEFCE(UK), NCSA(UIUC), KICP(U. Chicago), CCAPP(Ohio State), MIFPA(Texas A\&M), CNPQ, FAPERJ, FINEP (Brazil), MINECO(Spain), DFG(Germany) and the collaborating institutions in the Dark Energy Survey, which are Argonne Lab, UC Santa Cruz, University of Cambridge, CIEMATMadrid, University of Chicago, University College London, DES-Brazil Consortium, University of Edinburgh, ETH Zurich, Fermilab, University of Illinois, ICE (IEECCSIC), IFAE Barcelona, Lawrence Berkeley Lab, LMU Munchen and the associated Excellence Cluster Universe, University of Michigan, NOAO, University of Nottingham, Ohio State University, University of Pennsylvania, University of Portsmouth, SLAC National Lab, Stanford University, University of Sussex, and Texas A\&M University.

This research has made use of the NASA/IPAC Infrared Science Archive, which is funded by the National Aeronautics and Space Administration and operated by the California Institute of Technology.

\facilities{CTIO:Blanco (DECam), IRSA, SDSS}

\appendix

\section{Incompleteness Corrections for SDSS Data}\label{appendix:Incompleteness}
Using stars in the range $-21^{\circ}<\Lambda_{OCS}<-7^{\circ}$ and $10^{\circ}<\Lambda_{OCS}<21^{\circ}$ with $20.7<g_{corr}<21.7$, we fit the stellar density of F-turnoff stars $s(g_{corr})$ to a linear model:

    \begin{equation}\label{eq:s_bound}
        s(g_{corr})=m_sg_{corr} + b_s.
    \end{equation}
We exclude stars with $-7^{\circ}<\Lambda_{OCS}<10^{\circ}$ to avoid contamination from the Sagittarius Stream. As seen in Figure \ref{fig:g_corr_bin}, the line that fits the on-field is different from that of the off-field, so we must fit them separately. Binning over 20 $g_{corr}$ bins ($[g_1,g_2,...,g_{20}]$), we calculate the number of stars that fall within each bin, differentiating between the on-field and off-field. After normalizing these star counts, we use the ``curve fit'' algorithm from SciPy \citep{2020SciPy-NMeth} to calculate the slope ($m_{\rm on/off}$) and intercept ($b_{\rm on/off}$)  and their respective errors ($\delta m_{\rm on/off}$,$\delta b_{\rm on/off}$) for the on-field and off-field:

    \begin{equation}\label{eq:m_fits}
    \begin{aligned}
        m_{\rm on}\pm\delta m_{\rm on}=(8.16\pm1.62)\times10^{-3},\\
        m_{\rm off}\pm\delta m_{\rm off}=(4.91\pm1.68)\times10^{-3},
    \end{aligned}
    \end{equation}
    
    \begin{equation}\label{eq:b_fits}
    \begin{aligned}
        b_{\rm on}\pm\delta b_{\rm on}=-0.123\pm0.034,\\
        b_{\rm off}\pm\delta b_{\rm off}=-0.054\pm0.036.
    \end{aligned}
    \end{equation}

For each $i^{th}$ $\Lambda_{OCS}$ bin in the on-field subtending the $\Lambda_{OCS}$ range $\Lambda_{min,i}$ to $\Lambda_{max,i}$, we assume each completed bin in $(\Lambda,g_{corr})$-space has the form of a trapezoidal prism bounded by the fitted planes described in Equations \ref{eq:m_fits} and \ref{eq:b_fits}. This makes the total volume of such a bin equal to:

    \begin{equation}\label{eq:V_total}
        V_i=(21.4m_s + b_s)(\Lambda_{max,i}-\Lambda_{min,i}).
    \end{equation}
    
To calculate the volume of the bin that is actually filled, we slice this volume using the curve described in Equation \ref{eq:g_bound}. The infinitesimal volume $dv$ of a trapezoidal cross-section with width $d\Lambda$ is given by:

    \begin{equation}\label{eq:dV}
        dv=\frac{1}{2}(m_s(g_b(\Lambda)+20.7) + 2b_s)(g_b(\Lambda)-20.7)d\Lambda.
    \end{equation}
Substituting in $g_b(\Lambda)$, we find that the filled volume of the $i^{th}$ bin is:

    \begin{equation}\label{eq:v_filled}
    \begin{aligned}
        v_i=&\frac{m_sa_g^2}{10}\left(\Lambda_{max,i}^5-\Lambda_{min,i}^5\right) + \frac{m_sa_gb_g}{4}\left(\Lambda_{max,i}^4-\Lambda_{min,i}^4\right) + \frac{1}{3}\left(m_sa_gc_g + \frac{m_sb_g^2}{2} + b_sa_g\right)\left(\Lambda_{max,i}^3-\Lambda_{min,i}^3\right)\\
        +& \frac{1}{2}(m_sb_gc_g + b_sb_g)\left(\Lambda_{max,i}^2-\Lambda_{min,i}^2\right) + \left(m_s\frac{c_g^2-20.7^2}{2} + b_s(c_g-20.7)\right)\left(\Lambda_{max,i}-\Lambda_{min,i}\right).
    \end{aligned}
    \end{equation}
Dividing $v_i$ by $V_i$ gives us the filled ratio $k_i$ of the bin:

    \begin{equation}\label{eq:filled_frac}
        k_i=\frac{v_i}{V_i}.
    \end{equation}
Performing error propagation, we get the error $\delta_{k_i}$:

    \begin{equation}\label{eq:filled_frac_err}
    \begin{aligned}
        \delta_{k_i}^2=& \frac{1}{V_i^4}\left(\left(\frac{a_g^2}{10}\left(\Lambda_{max,i}^5-\Lambda_{min,i}^5\right) + \frac{a_gb_g}{4}\left(\Lambda_{max,i}^4-\Lambda_{min,i}^4\right) + \frac{1}{3}\left(a_gc_g + \frac{b_g^2}{2}\right)\left(\Lambda_{max,i}^3-\Lambda_{min,i}^3\right)\right.\right.\\
        +&\left.\left. \frac{b_gc_g}{2}\left(\Lambda_{max,i}^2-\Lambda_{min,i}^2\right) + \frac{c_g^2-20.7^2}{2}\left(\Lambda_{max,i}-\Lambda_{min,i}\right)\right)V_i - 21.4\left(\Lambda_{max,i}-\Lambda_{min,i}\right)v_i\right)^2\delta_m^2\\
        +& \frac{1}{V_i^4}\left(\left(\frac{a_g}{3}\left(\Lambda_{max,i}^3-\Lambda_{min,i}^3\right) + \frac{b_g}{2}\left(\Lambda_{max,i}^2-\Lambda_{min,i}^2\right) + (c_g-20.7)\left(\Lambda_{max,i}-\Lambda_{min,i}\right)\right)V_i \right.\\
        -&\left. \left(\Lambda_{max,i}-\Lambda_{min,i}\right)v_i\right)^2\delta_s^2
    \end{aligned}
    \end{equation}
    
After calculating the filled ratio, we approximate the corrected star count ($N'_i$) and errors ($\sigma_{N'_i}$) using the following formulas:

    \begin{equation}
        N'_i = \frac{N_i}{k_i},
    \end{equation}
    \begin{equation}
        \sigma_{N'_i} = \frac{1}{k_i}\sqrt{N_i + \left(\frac{N_i}{k_i}\right)^2\delta_{k_i}^2}.
    \end{equation}
This algorithm is repeated for each bin in the off-field as well.

\section{Calculating Mass per Turnoff Star Using Isochrones}\label{appendix:Isochrone}
As a sanity check for our previous calculation, we recalculate the baryonic mass per F-turnoff star using theoretical isochrones of NGC 5053. To perform this calculation, we first need to know three things about NGC 5053: its age, its metallicity ([Fe/H]), and its alpha abundance ([$\alpha$/Fe]). From the literature, we find NGC 5053's age to be $12.5\pm2.0$ Gyrs \citep{ngc5053_age}, its [Fe/H] to be -2.27 dex \citep[][2010 edition]{ngc5053_metal}, and its [$\alpha$/Fe] to be 0.2 \citep{ngc5053_alpha}. We use isochrone data from the Dartmouth Stellar Evolution Database \citep[DSED;][]{dartmouth_database}, selecting the isochrones that best fit the globular cluster's age and chemical abundances. For our calculations, we use the isochrone with [Fe/H]$=-2.49$ dex as it was the closest metallicity. However, we will later demonstrate that this deviation does not greatly impact our result or errors by redoing the calculation assuming [Fe/H]$=-1.98$ dex.

The initial mass function (IMF) we implement comes from \cite{IMF_profile}, and has the form:

\begin{equation}
    \varepsilon(m)=
    \begin{cases}
    Am^{-\alpha_0}\indent& 0.01\leq m<0.08\\
    A{(0.08)}^{\alpha_1-\alpha_0}m^{-\alpha_1}& 0.08\leq m<0.5\\
    A{(0.5)}^{\alpha_2-\alpha_1}{(0.08)}^{\alpha_1-\alpha_0}m^{-\alpha_2}& 0.5\leq m<1.0\\
    A{(1.0)}^{\alpha_3-\alpha_2}{(0.5)}^{\alpha_2-\alpha_1}{(0.08)}^{\alpha_1-\alpha_0}m^{-\alpha_3}& m>1.0,\\
    \end{cases}
\end{equation}
where $m$ is the mass of the initial progenitor star in solar masses, $A$ is the normalization constant, $\alpha_0 = 0.3\pm0.4$, $\alpha_1 = 1.3\pm0.3$, $\alpha_2 = 2.3\pm0.1$, and $\alpha_3 = 2.3\pm0.2$. The uncertainties in these powers are one-standard-deviation errors.

The formula we use to calculate the stellar mass per turnoff star is fairly straightforward. We take the total stellar mass within the cluster and divide it by the number of F-turnoff stars we find in our isochrone:

\begin{equation}
    m_{FT} = \frac{M_{stars}}{N_{FT}} = \frac{\int m\varepsilon(m)dm}{\int \varepsilon(m)dm}.
\end{equation}
Note that since the IMF is in both our numerator and denominator, the normalization constant $A$ cancels out of our formula, making the normalization we select arbitrary. The easiest quantity to calculate is $N_{FT}$, the number of F-turnoff stars in our isochrone. We say a star is an F-turnoff star if its $(g-i)_0$ color falls between 0.12 and 0.47, and its $g_0$ magnitude within the $g_0$ F-Turnoff Range of the isochrone, using the same convention developed in Section \ref{sec:est_mass} to define the range.

We define a function we call the F-check function ($F_C(m)$) which outputs 1 if the input initial mass $m$ falls within our F-turnoff range as a result of our isochrone model and 0 otherwise:

\begin{equation}
    F_C(m)=
    \begin{cases}
    1\indent&\text{m is F-turnoff star}\\
    0&\text{otherwise}.\\
    \end{cases}
\end{equation}
Given this function and a list of initial masses from the isochrone model ($[m_1,m_2,\dots,m_N]$), the formula for counting the number of F-turnoff stars becomes straightforward:

\begin{equation}
    N_{FT}=\int F_C(m)\varepsilon(m)dm \simeq \sum_{i=1}^{N}F_C(m_i)\varepsilon(m_i)\Delta m_i,
\end{equation}
where $\Delta m_i$ is the width of the i$^{th}$ mass bin defined below:

\begin{equation}
    \Delta m_i=
    \begin{cases}
    m_2-m_1\indent&i=1\\
    m_N-m_{N-1}&i=N\\
    \frac{m_{i+1}-m_{i-1}}{2}&\text{otherwise}.\\
    \end{cases}
\end{equation}

Calculating the total stellar mass, on the other hand, is a bit more complicated. For stars whose initial mass is larger than the largest mass in the isochrone model ($m > m_N$), their current mass is only but a small fraction of their original mass. This is because such stars have turned into white dwarfs, neutron stars, or black holes, ejecting most of their original mass through their planetary nebulae or supernovae. This ejected mass does not necessarily disappear from the globular cluster, but could be accelerated with enough energy to push it past the cluster's escape velocity. We know the mass of NGC 5053 is $(5.37\pm1.32)\times10^4M_\odot$, and from our previous fit, we find the scale Plummer radius of NGC 5053 to be $11.7\pm0.5$ pc. Performing a simple back-of-the-envelope calculation of the cluster's highest escape velocity $(v_{e,max} = \sqrt{\frac{2GM}{a}})$ gives an escape velocity of $\sim6.4$ km s$^{-1}$. Since planetary nebulae expand from their center stars with speeds between 20 and 40 km s$^{-1}$ \citep{schonberner2014}, and the shocks of supernovae reach speeds of several thousand km s$^{-1}$ \citep{hovey2015,hovey2016}, it is safe to assume that all ejecta of stellar transitions is not gravitationally bound by the cluster and can effectively be ignored in the mass calculation.

Under these assumptions, the total stellar mass can be calculated using the following formula:

\begin{equation}
    M_{stars} = \int_{0.01}^{m_N+\frac{\Delta m_N}{2}} m \varepsilon(m)dm + \int_{m_N+\frac{\Delta m_N}{2}}^{100.0} M_R(m,[Fe/H]) \varepsilon(m)dm,
\end{equation}
where $M_R(m,[Fe/H])$ is the remnant mass function, a function that outputs the stellar remnant mass of a star given its initial progenitor mass. We use the models fitted in \cite{cummings2018} to calculate the remnant mass of white dwarfs and the models from \cite{fryer2012} to account for the masses in neutron stars and black holes. We cut off our mass calculation at an initial mass of 100 solar masses as the remnant masses past that point become more uncertain due to mass loss and pair-instability supernovae which leave behind no remnant \citep{fryer2012}. Combining these models, we find the mass of the remnant in solar masses is given by the following formula:

\begin{equation}
    M_R(m,[Fe/H])=
    \begin{cases}
    (0.080\pm0.016)m+(0.489\pm0.030)\indent&m<2.85\\
    (0.187\pm0.061)m+(0.184\pm0.199)&2.85\leq m<3.60\\
    (0.107\pm0.016)m+(0.471\pm0.077)&3.60\leq m<9.0+0.9[Fe/H]\\
    1.36&9.0+0.9[Fe/H]\leq m<11.0\\
    1.1+0.2e^{(m-11.0)/4.0}-(2.0+10^{[Fe/H]})e^{0.4(m-26.0)}&11.0\leq m<30.0\\
    \Theta(m,[Fe/H])&30.0\leq m<50.0\\
    max(1.8+0.04(90-m),\Theta(m,[Fe/H]))&50.0\leq m<90.0\\
    max(1.8+\log_{10}(m-89),\Theta(m,[Fe/H]))&m>90.0,\\
    \end{cases}
\end{equation}
where

\begin{equation}
    \Theta(m,[Fe/H])=min(33.35+(4.75+1.25\times10^{[Fe/H]})(m-34),m-10^{[Fe/H]/2}(1.3m-18.35)).
\end{equation}

To propagate the errors in this calculation, we recompute the stellar mass per turnoff star several times, changing the parameters with errors by their respective error. We have 11 quantities with errors in this calculation: the age, the four powers in the IMF, and the six parameters from the white dwarf remnant formula derived by \cite{cummings2018}. For each parameter, we either add its error, subtract its error, or leave the parameter unchanged. We calculate the mass per turnoff star for each permutation ($3^{11}$ total) and treat each one as a separate data point. We then take the average value and calculate the standard deviation of all the points as the error. Using this method, we find that the stellar mass per F-turnoff star we expect to measure in NGC 5053 to be $13.5\pm2.7 M_{\odot}$ per F-turnoff star. Using an isochrone with [Fe/H]=-1.98 dex, we get an answer of $14.9\pm3.5 M_{\odot}$ per turnoff star. Both of these numbers are exceptionally close to the value we measure in Section \ref{sec:est_mass} and well within the expected errors.

\clearpage
\nocite{*}
\bibliographystyle{aasjournal}
\bibliography{research}

\end{document}